%% file: xray_corr_accepted.tex
\documentclass[iop]{emulateapj}
\usepackage{graphicx}
\usepackage{amssymb}
\usepackage{amsmath}
\usepackage{epsfig}
\usepackage{natbib}
\usepackage{booktabs}
\usepackage[normalem]{ulem}

\newcommand{\myemail}{judith.racusin@nasa.gov}
\newcommand{\swift}{{\it Swift }}

\slugcomment{{\sc Accepted for publication in ApJ}}

\shorttitle{X-ray Average Decay - Luminosity Correlation}
\shortauthors{Racusin et al.}

\begin{document}

\title{A Correlation Between the Intrinsic Brightness and Average
  Decay Rate of Gamma-ray Burst X-ray Afterglow Light Curves}

\author{J. L. Racusin\altaffilmark{1,2},
  S. R. Oates\altaffilmark{3,4}, M. de Pasquale\altaffilmark{4}, and
  D. Kocevski\altaffilmark{1}} 
\altaffiltext{1}{NASA's Goddard Space
  Flight Center, Code 661, Greenbelt, MD} \altaffiltext{2}{\myemail}
\altaffiltext{3}{Instituto de Astrofsica de Andalu\'{c}ıa (IAA-CSIC),
  Glorieta de la Astronom\'{i}a s/n, E-18008, Granada, Spain}
\altaffiltext{4}{Mullard Space Science Laboratory, University College
  London, Holmbury St. Mary, Dorking, Surrey RH5 6NT}

\begin{abstract}
  We present a correlation between the average temporal decay
  ($\alpha_{X,avg,>200s}$) and early-time luminosity ($L_{X,200s}$) of
  X-ray afterglows of gamma-ray bursts as observed by {\it Swift}-XRT.
  Both quantities are measured relative to a rest frame time of 200 s
  after the $\gamma$-ray trigger.  The luminosity average decay
  correlation does not depend on specific temporal behavior and
  contains one scale independent quantity minimizing the role of
  selection effects.  This is a complementary correlation to that
  discovered by \cite{2012MNRAS.426L..86O} in the optical light curves
  observed by {\it Swift}-UVOT.  The correlation indicates that on
  average, more luminous X-ray afterglows decay faster than less
  luminous ones, indicating some relative mechanism for energy
  dissipation.  The X-ray and optical correlations are entirely
  consistent once corrections are applied and contamination is
  removed.  We explore the possible biases introduced by different
  light curve morphologies and observational selection effects, and
  how either geometrical effects or intrinsic properties of the
  central engine and jet could explain the observed correlation.

\end{abstract}

\section{Introduction}

Observations and understanding of Gamma-ray Bursts (GRBs) has changed
dramatically over the last decade with NASA's \swift Gamma-ray Burst
Explorer mission \citep{gehrels04}. \swift has provided a vast
database of both prompt emission observations with the Burst Alert
Telescope (BAT; \citealt{barthelmy05}), and early ($T_{start}<T_0+100\
\textrm{s}$) afterglow observations with the X-ray
Telescope (XRT; \citealt{burrows05}) and with the
Ultraviolet Optical Telescope (UVOT; \citealt{roming05}).
Correlations between observable parameters in GRBs are plentiful but
at present, most are not well understood.  The relationships between
the physical and geometrical effects needs to be separated in order to
determine the true nature of GRB progenitors and jet physics of the
central engine.

\cite{2012MNRAS.426L..86O,oates15} presented a correlation between the early-time
($t_{rest,200\textrm{s}}=t_{obs}/(1+z)=200~\textrm{s}$, hereafter
abbreviated $t_{200\textrm{s}}$) luminosity ($L_{O,200\textrm{s}}$) of
a GRB optical afterglow light curve and the average decay rate of that
light curve ($\alpha_{O,avg,>200\textrm{s}}$) from $t_{200\textrm{s}}$
onward, measured by fitting a power law, with a linear relationship
$\alpha_{O,avg,>200\textrm{s}}=(-0.28\pm0.04)\textrm{log}\
L_{O,200\textrm{s}}+(7.72\pm1.31)$,
with a Spearman rank correlation coefficient (hereafter $R_{sp}$) of
-0.58 and a null hypothesis of $1.90\times10^{-5}$.  This qualitatively implies
that initially brighter GRB afterglows decay faster and fainter afterglows decay
slower.  In this paper, we aim to search for this same
correlation with X-ray afterglow observations.  X-ray light curves
have been shown to typically contain more complex structure than
optical light curves \citep{panaitescu07,liang08,evans09}, as the morphology
tends to be the convolution of the prompt emission internal
dissipation mechanisms and the afterglow forward shocks.  Therefore,
the average decay rate depends on which light curve segments are
included, which in turn depends on the assumptions on the physical
origins of these segments.

Other trends and correlations have been demonstrated for X-ray
afterglows including some evidence for clustering in luminosity with
the brighter afterglows decaying faster than the fainter ones
\citep{boer00,gendre05}, though that clustering has disappeared with
larger samples \citep{gendre08,bardho15}; a correlation between the
luminosity and time of the end of the plateau
\citep{dainotti10,dainotti13,sultana13}, and correlations with prompt
emission parameters \citep{gehrels08,dainotti11,davanzo12,margutti13}.

The data reduction and analysis including sample selection and light
curve fitting, will be discussed in \S\ref{sec:analysis}, the results
of the correlation tests in \S\ref{sec:results}, observational biases,
other X-ray afterglow correlations, and possible physical origin are
discussed in \S\ref{sec:disc}, and conclusions in \S\ref{sec:conc}.
Throughout this work, the convention 
$F(t)\sim t^{-\alpha}\nu^{-\beta}$ is used, where $\alpha$ is the
temporal power-law index and $\beta$ is the spectral energy index.
Note that \cite{2012MNRAS.426L..86O,oates15} use the negative of this convention.
All errors are $68\%$ confidence unless stated otherwise, and we
assume cosmological parameters of $H_0=70\ \textrm{km}\
\textrm{s}^{-1}\ \textrm{Mpc}^{-1}$, $\Omega_\Lambda=0.7$ and $\Omega_m=0.3$.

\section{Data Reduction and Analysis} \label{sec:analysis}

\subsection{Sample Selection}

All {\it Swift}-BAT discovered GRBs, with X-ray afterglows detected by
\swift-XRT and measured redshifts, discovered between December 2004
and March 2014 are included in the following analysis.  Only those
bursts with redshifts can be used, because the correlation parameters
depend on intrinsic luminosity at the same relative redshift-scaled
time.  We include only those X-ray afterglows with at least 3 light
curve bins ($\gtrsim60$ counts), such that we have sufficient
statistics to constrain the temporal and spectral fits, and $T_{90}$
measurements are available in the BAT catalogs
\citep{sakamoto08,sakamoto11,lien15}.  There are 280 GRBs in our
sample that fit this criteria.  As described in \S\ref{sec:shortlong},
we compare the correlation for both short and long duration GRBs, but
primarily focus on long bursts, as short bursts to a large extent have
different light curve morphology, environments, and energetics.  The
early average decay - luminosity correlation focuses on early-time
flux estimates, therefore we also exclude any bursts for which
observations started more than a factor of 2 after
$t_{200\textrm{s}}$, such that extrapolation over long timescales
could be inaccurate.  The final sample includes 237 long GRBs (9
short), 47 of which also appear in the
\cite{2012MNRAS.426L..86O,oates15} sample for the UVOT correlation
(sample only extends through 2010).  The redshift measurements come
from a convolution of
databases\footnote{\url{http://www.mpe.mpg.de/~jcg/grbgen.html}}$^,$\footnote{\url{http://www.astro.caltech.edu/grbox/grbox.php}}
and the literature, and are listed in Table \ref{table:everything}.


\subsection{Light Curve Fitting} \label{sec:lcs}

All light curves were retrieved from the University of Leicester
\swift XRT Team GRB repository \citep{evans07,evans09}.  The
count rate light curves were converted to 0.3-10 keV flux using a
single conversion factor from the automated repository spectral fits
to the photon counting (PC) mode data.  XRT switches to PC mode typically after
the episodes of significant spectral evolution during flares and steep
decay segments, yet while the afterglow is bright enough to accumulate sufficient
statistics for a well-measured spectrum.  However, all light curve
fitting, as described in this section, was done in the count rate
domain.  

The flux light curves were converted to 1 keV flux density using the
spectral index from the automated PC fits, and then to intrinsic
luminosity, and k-corrected using the
following formula:
\begin{equation}
L(t)=F_\nu(t)\times 4\pi D^2 (1+z)^{(\beta-1)},
\end{equation}
where $L(t)$ is the luminosity at any given time, $F_\nu(t)$ is the 1 keV
flux density, $D$ is the luminosity distance, and $z$ is the redshift.

The BAT trigger time, $T_0$, is generally used as the zero point of
light curve analysis.  However, $T_0$ for each GRB is a quantity
determined by one of hundreds of onboard trigger criteria, and as
such, is an empirical quantity, not a physical measure of the burst
onset.  While it typically does well as an indication of onset,
occasionally emission is clearly visible prior to $T_0$.  The standard
measurement of GRB duration - the time in which $90\%$ of the fluence
(time-integrated flux) is emitted ($T_{90}$) - is at least a
characterization of the measured emission, though can be significantly
flawed in cases with multiple emission episodes with quiescent
emission in between, and varies with the relative level of signal and
noise.  In the BAT automated and catalog temporal analyses
\citep{sakamoto08,sakamoto11,lien15}, $T_{90}$ is measured between
$T_{05}$ and $T_{95}$, which represents the $5\%$ and $95\%$ limits on
the BAT GRB emission.  In this paper, we use the time $T_{05}$ instead
of $T_0$ as the light curve zero point.  In $64\%$ of GRBs in our
sample, $T_0$ and $T_{05}$ differ by less than 5 seconds, and less
than one minute in $96\%$ of the sample, with extreme cases (likely
due to \swift slewing at the time of emission onset or another
instrumental delay) deviating by as much as 200-300 seconds.  The
average decay rate is typically affected by the shift of $T_0$ with a
change in slope $<0.1$.  All BAT parameters were taken from the BAT
catalogs \citep{sakamoto08,sakamoto11,lien15}.



\begin{figure}
\includegraphics[width=0.5\textwidth,trim=80 10 0 100,clip=true]{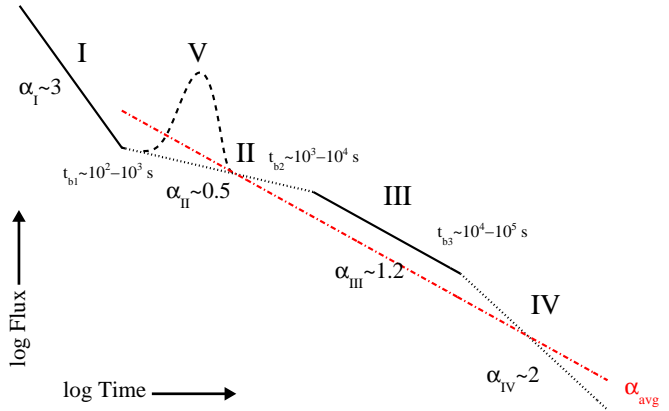}
\caption{Morphology of canonical GRB X-ray afterglow
  \citep{zhang06,nousek06} and comparison to average afterglow.
  Segments are generally attributed to: I) steep decay due to
  curvature effect of delayed prompt emission; II) plateau due to some
  form of continued energy injection; III) normal foward shock decay
  of the afterglow; IV) the post-jet break decay; and V) flares
  that are seen in $\sim1/2$ of all GRBs.  The measurements of the
  average decay rate are described in \S\ref{sec:lcs}.  Slopes and
  break times have broad distributions (see
  \citealt{racusin09,evans09}.)  \label{fig:canon}}
\end{figure}

GRB X-ray afterglow light curves demonstrate complex and highly varied
structure, and typically contain up to 5 components (Figure
\ref{fig:canon}): I) a steep decay lasting tens to hundreds of seconds
likely related to the tail of the prompt emission generally attributed
to the curvature effect; II) a plateau lasting hundreds to thousands of
seconds which is generally attributed to some sort of energy injection
boosting the forward shock emission; III) the normal forward shock
power-law decay; IV) a post-jet break decay; and V) in some cases flares
which can occur during any phase and be either single emission
episodes or in multiple \citep{zhang06,nousek06}, and an occasional
flat segment prior to the steep decay labeled as 0, which is part of
the prompt emission.  While the majority
of bursts can be characterized as displaying some of the canonical
components, very few rarely display them all, and significant
deviations such as single power-law behavior is common
\citep{racusin09,evans09}. 

In our previous works \citep{racusin09,racusin11}, we have
characterized the best-fit power-laws to the full sample of X-ray
afterglows.  We have since added Gaussians to characterize X-ray
flares, rather than removing the interval by eye from the fitted data.
Our process still requires user input, and is only semi-automated.
The power law fits with up to 5 breaks are used as a way to
differentiate the internal (steep decay/flares) versus external
(plateaus, normal, post-jet break) shock portions of the light curves.
These fits to all segments I-V will be hereafter referred to as
``best-fit (power-law) models'' (i.e. Figure \ref{fig:canon}). We
measure an early characteristic 
luminosity, $L_{X,200\textrm{s}}$ at 1 keV, by extrapolating or interpolating
the luminosity at $t_{200}$ from the best fit (power-law) models
depending on when the observations began.
The temporal decay index from the best fit (power-law) models at $t_{200s}$
is referred to as $\alpha_{X,fit,200s}$.

\begin{figure}
\includegraphics[width=0.45\textwidth,trim=120 20 50 30,clip=true]{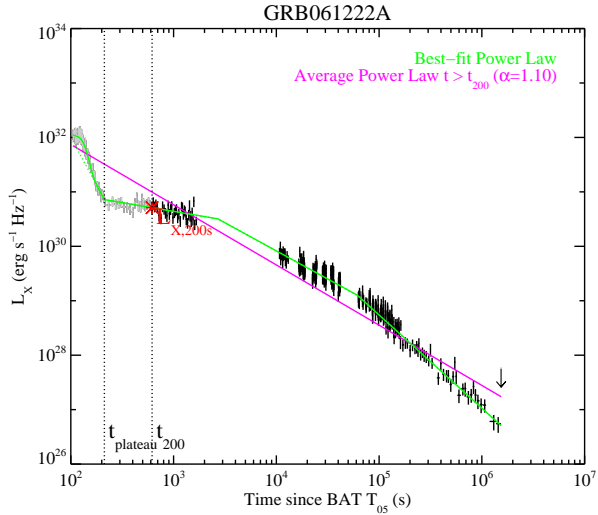}
\caption{Light curve and and fits to GRB
  061222A, an example X-ray afterglow that displays the canonical
  best-fit power law (PL) segments (green), and a single average power
  law fit (magenta) demonstrating our measure of average decay
  rate ($\alpha_{X,avg,>200\textrm{s}}$).  The luminosity
  $L_{X,200\textrm{s}}$ is extracted from the best fit
    (power-law) models at $t_{200\textrm{s}}$.   The points 
  excluded from the average power-law models are in gray.
\label{fig:lc}}
\end{figure}

We attempt to characterize average behavior (hereafter referred to as
``average power-law models''), not focusing on the details that affect the
individual slopes such as the density profile of the circumburst
environment, location and evolution of the cooling break in the
Synchrotron spectrum \citep{sari98}, and the microphysical parameters
(electric and magnetic field contribution, density of environment,
radiative efficiency).  Therefore, the simplest way to measure the
average decay rate ($\alpha_{X,avg,>200\textrm{s}}$) is to fit a
simple power-law to the complex light curve of the X-ray afterglow
(Figure \ref{fig:lc}), after some specific rest frame time, for which
we use $t_{200\textrm{s}}=200 \times (1+z)$.  While this is most
likely a very poor fit statistically, it is a adequate characterization
of the average behavior.   

Complications affect X-ray light curves more so than their
optical correlaries including the tail of the prompt emission, and
early- and late-time flares.  Therefore in \S\ref{sec:results} and
Figure \ref{fig:decaylum_initial}, we will demonstrate 
the luminosity-average decay correlation without performing any corrections
to the light curves.   In the following, we demonstrate step-by-step how our
correlation is improved if we exclude these additional components from
the measures of $L_{X,200\textrm{s}}$ and $\alpha_{X,avg,>200\textrm{s}}$.

\begin{figure}
\includegraphics[width=0.45\textwidth,trim=120 20 50 30,clip=true]{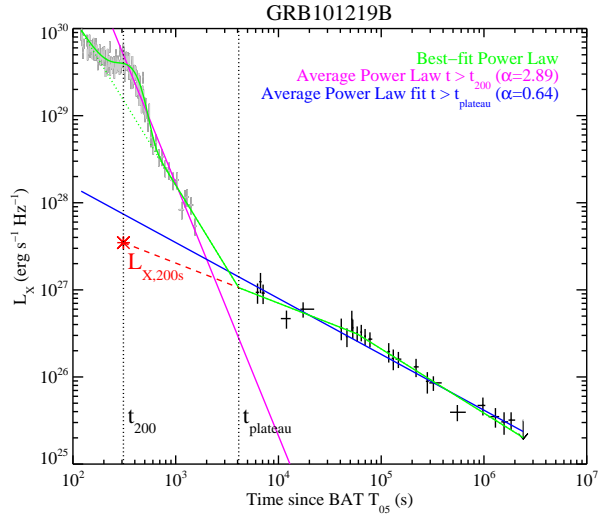}
\caption{Fits to the X-ray light curve of GRB 101219B, for which the
  uncorrected average decay (magenta) is contaminated by
  the steep decay segment (Segment I).  As 
  described in the text, we wish to characterize only the external shock afterglow,
  and therefore fit the average decay after the transition from steep
  to shallow decay ($t>t_{plateau}$, Segments II-IV, blue), and
    extrapolate the best fit (power-law) models back to $t_{200\textrm{s}}$ to measure 
  $L_{X,200\textrm{s}}$ (red).  The initial average 
  power law fit for $t>t_{200\textrm{s}}$ (magenta) is much steeper than the
  corrected average power law fit to $t>t_{plateau}$ (blue).  The points
  excluded from the average power-law models are in gray.
  \label{fig:steep_contam}}
\end{figure}

The average power-law model is dominated by the portion of the light
curve with the best statistics - the early bright part (often
  Segment I), which is also
most likely to be related to the prompt emission, not the afterglow
\citep{zhang07}.  For those cases where the average decay is dominated
by the steep decay (i.e. it ends after $t_{200\textrm{s}}$), we
correct their average decay fits by including only the data after the
end of the steep decay (Segments II-IV identified through the
best fit (power-law) models, see \S\ref{sec:steepdecay} for
details).  In these cases $L_{X,200\textrm{s}}$ will also be
dominated by the prompt emission.  To obtain a luminosity
representative of the afterglow at that time we extrapolate Segment II of the 
best fit (power-law) models back to $t_{200\textrm{s}}$.  Figure 
\ref{fig:steep_contam} demonstrates an example case that we correct
for steep decay contamination in the average decay and luminosity.

Cases where there are flares after $t_{200\textrm{s}}$ (or the start
of the plateau, if later), may also be significantly biased by the
bright light curve points during flares that dominate the emission.
We test the impact of flares on the average decay-luminosity
correlation in \S\ref{sec:flares}, and alleviate this issue by
excluding all the of the light curve points that contain flaring
intervals when fitting the average power-law model.

\subsection{Correlation Analysis\label{sec:corr_anal}}

In order to determine the strength and significance of each
correlation, we perform a linear regression analysis using the IDL
astrolib\footnote{\url{http://idlastro.gsfc.nasa.gov/}} routine {\it
  fitexy}.  The advantage of this tool is that it utilizes errors on
both parameters, and is reverseable if one switches the axes of the
input parameters.  Therefore, it does not require a dependent variable.  From
this regression analysis, we measure the slope of the relation and a
constant offset (Table \ref{table:corr}).  Although {\it fitexy}
accounts for the errors on both parameters, it does not address the
intrinsic dispersion.  Consequently, we find a reduced $\chi^2>>1$, and
therefore the covariance errors are not reliable.  To overcome this
limitation, we determine the errors via a Monte Carlo Bootstrap
calculation.  For $10^4$ trials, we randomly selected a subset of our
data sample that is the same size as our data sample, allowing for
possible duplicated values (i.e. sampling with replacement), and redo
the linear regression.  The resulting distributions of slope and
constant offset for each trial, allow us to extract the $1\sigma$
errors on each parameter.

We determine the strength of the correlation by measuring a Spearman
rank coefficient ($R_{sp}$), and its corresponding null hypothesis
probability ($p$)
using the IDL tool {\it r\_correlate}.  We also test the dependence on
the correlation between $L_{X,200\textrm{s}}$ and
$\alpha_{X,avg,>200\textrm{s}}$ with redshift, a parameter with some
dependence in both quantities.  To do this, we conduct a partial
Spearman rank correlation analysis.  The correlation remains
significant when accounting for redshift (Table \ref{table:corr}),
demonstrating that we are not simply correlating redshift with itself.


\section{Results} \label{sec:results}

The following section describes the resulting luminosity - average
decay correlation, and how we reduce the scatter, optimize the
correlation, and investigate its origin.  The correlation parameters
for each iteration and subset are summarized in Table \ref{table:corr}.

\begin{figure}[!t]
\includegraphics[width=0.5\textwidth,clip=true,trim=0 50 0 50]{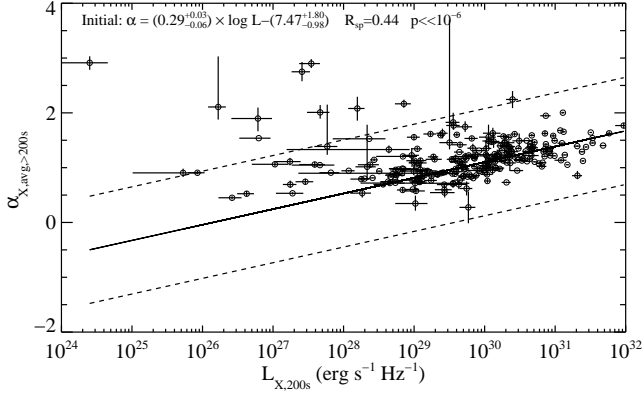}
\caption{Average decay - luminosity correlation using the full sample with
  no optimizations, corrections or filtering.  There is significant
  scatter at low luminosities which will be addressed in
  \S\ref{sec:results}.  The best fit regression parameters and their
  $1\sigma$ errors, and the Spearman rank coefficient and associated
  probability are listed on the plot and in Table \ref{table:corr}.
  The solid line indicates the best fit regression, and dashed lines indicates
  the $2 \sigma$ deviation.\label{fig:decaylum_initial}} 
\end{figure}


Similar to the correlation in \cite{2012MNRAS.426L..86O}, we see
evidence for a correlation (Figure \ref{fig:decaylum_initial}) in our
full sample, even without correcting for the steep decay
contamination.  Significant scatter is present in the relation, due to
both the steep decay contamination discussed in \S\ref{sec:steepdecay}, and
fundamental differences in sub-samples (e.g. long/short duration, \S\ref{sec:shortlong}).
The next section describes optimizations and
tests for light curve morphological features that may influence the
correlation and apparent scatter.


The presence of features in the afterglow light curves (plateaus,
flares, etc.) could potentially influence or even be the root cause of
the correlations.  We test these effects, and other observational
biases by splitting the sample and reproducing the same analyses in
the following.  Through these efforts we are able to
reduce some of the sources of the scatter in the relation.

\subsection{Steep Decay Contamination\label{sec:steepdecay}}

\begin{figure}[!h]
\includegraphics[width=0.5\textwidth,clip=true,trim=0 50 0 50]{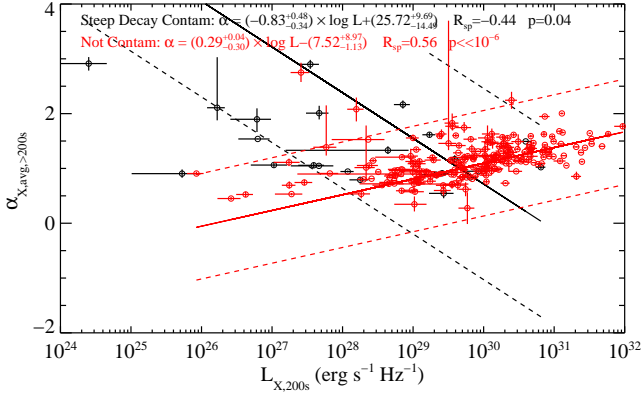}
\caption{Average decay - luminosity correlation using the full sample
  with no optimizations, corrections or filtering.  The sample is
  split between those that require a steep decay contamination
  correction (but have not yet received it), and those that do not
  require it.  The sample that does not require the correction shows a
  significant correlation, while the sample that does require the
  correction shows no evidence for a significant correlation.  The
  solid line indicates the best fit regression, and the dashed lines
  indicates the $2 \sigma$ deviation.\label{fig:decaylum_corr}}
\end{figure}

As demonstrated in Figure \ref{fig:steep_contam}, GRBs for which the
steep decay continues past $t_{200s}$, have average light curve fits
contaminated by that steep decay which is a distinct component from
the afterglow in both temporal and spectral morphology.  This affects
both of our measurements of $\alpha_{X,avg,>200\textrm{s}}$ and
$L_{X,200\textrm{s}}$. Of the 246
GRBs in our sample, 23 suffer from this contamination.  In those
cases, we fit the average light curve over the interval from the start
of the plateau to the end of the light curve.  A comparison between
those bursts that do and do not require the steep decay correction show
significant differences in the average decay-luminosity correlation
using the uncorrected decay rates (Figure \ref{fig:decaylum_corr}).
All correlations presented in the rest of this paper apply the steep
decay contamination correction to those 23 GRBs.

\LTcapwidth=\textwidth
\begin{longtable*}{|p{1.6cm}|c|c|c|c|c|c|c|c|c|}
\hline
Sample & \multicolumn{2}{c|}{Parameters} & 
Spearman & Null & Partial  &
Null & \multicolumn{2}{c|}{Best fit linear regression} & Number \\
& x-axis & y-axis
 & Rank & Hypothesis & 
Spearman  & Hypothesis & Slope & Constant & in \\
& & 
& & & Rank & & & & Sample \\
\hline
\hline
\input{results_table}
\caption{Regression analysis results and correlation statistics for
  each subsample and correlation. The partial Spearman rank
  coefficient tests the dependence on GRB redshift.  The significance
  columns apply to the regular or partial Spearman rank coefficient
  to the left of that column. Note that the subscripts ``avg'' and
  ``fit'' refer to the slope of the average temporal power law decay
  and the individual segment slopes of the best fit (power-law)
    models, respectively. \label{table:corr}} 
\end{longtable*}

\begin{figure}[!h]
\includegraphics[width=0.5\textwidth,clip=true,trim=0 50 0 50]{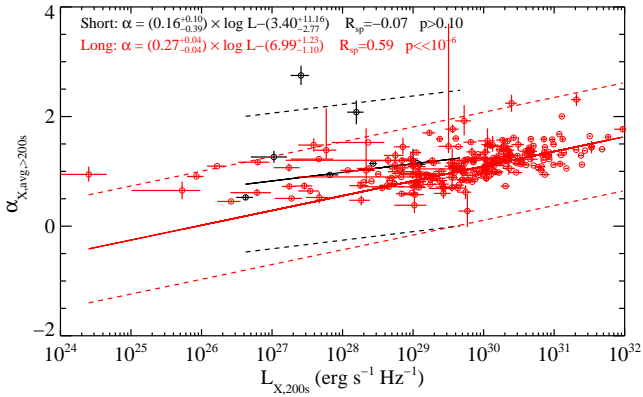}
\caption{Average decay - luminosity correlation using the sample
  including steep decay corrections, and split by short ($T_{90}< 2
  s$) and long ($T_{90}> 2 s$) prompt emission durations.  The sample
  of short GRBs is small and does not show any evidence of a
  significant correlation.  Short GRBs are removed for subsequent
  analyses.  The solid line indicates the best fit regression, and the
  dashed lines indicates the $2 \sigma$ deviation. \label{fig:decaylum_dur}}
\end{figure}

\subsection{Short versus Long Duration GRBs\label{sec:shortlong}}

Short duration GRBs display both similarities and differences to long
GRBs \citep{barthelmy05b,berger14}, including: prompt emission spectral lags
\citep{norris06,norris11}, redshift distributions \citep{guetta06},
prompt emission energetics \citep{amati02,ghirlanda09}, prompt
emission light curves \citep{nakar02}, prompt emission spectral
properties  \citep{goldstein12}, environments and host galaxies
\citep{berger07,troja08,fong13}, and afterglow properties
\citep{kann11,racusin09,nysewander09}.  
While approximately half of all short GRB
X-ray afterglows show long-lasting emission, the other half fade below
the level of detectability within the first few thousand seconds, with
on average steeper slopes than long GRBs, likely reflecting the lower
density circumstellar environments).  The so-called
``naked GRBs'' \citep{page06,perley09,hascoet11} may occur in such low density
environments, in which they do not display forward shock emission, or
the light curve decays below the detectability threshold during the steep decay phase.

We test for differences in the average decay-luminosity correlation
between short and long duration GRBs (Figure \ref{fig:decaylum_dur}),
finding that long duration bursts are significantly correlated
($R_{sp}=0.59$, $p=1.01\times10^{-23}$) in the average decay-luminosity
domain, and short bursts demonstrate no significant correlation
($R_{sp}=-0.07$, $p=0.86$) and in fact include two of the
points with the most deviant scatter.  This suggests that the
correlation is related to some difference between short and long GRBs
in their the afterglow properties, be it environment or jet dynamics.
For all further tests of the X-ray average decay-luminosity
correlation, we exclude short GRBs.

\subsection{X-ray Flares}\label{sec:flares}
\begin{figure}
\includegraphics[width=0.5\textwidth,clip=true,trim=0 50 0 50]{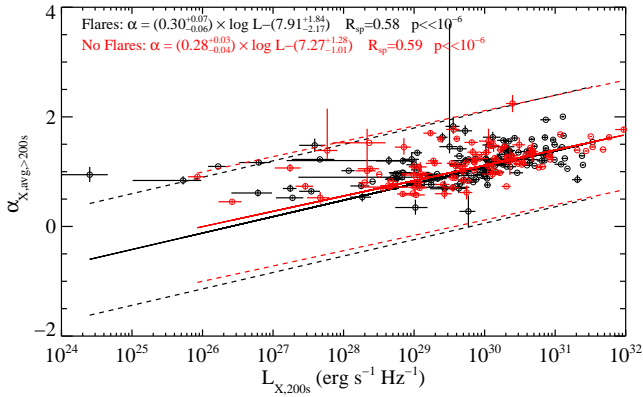}
\caption{Average decay - luminosity correlation using the sample
  including steep decay corrections, those only of long duration, and
  split into those with and without X-ray flares anywhere during the
  X-ray afterglow.  The correlation strengths and regression lines are
  consistent between the subsamples, with slightly more scatter in the
  sample with flares.  We remove the time intervals of significant
  flaring and refit the average decay for the rest of this analysis.
  The solid line indicates the best fit regression, and the dashed lines indicates
  the $2 \sigma$ deviation. 
  \label{fig:decaylum_flares}}
\end{figure}

X-ray flares have been extensively studied using early afterglow
observations from XRT, and have been shown to have an internal rather
than external shock origin
\citep{falcone07,chincarini07,chincarini10,kocevski07,margutti10}.
X-ray flares can occur at a variety of timescales and either as
single or multiple pulses.  As a potential source of contamination for
the average decay-luminosity correlation, we separate those afterglows
with X-ray flares (without removing flaring intervals) and those
without flares (Figure
\ref{fig:decaylum_flares}), and find that the two samples show very
similar correlation strengths and slopes, but with slightly more
scatter in the sample with flares.  Once we removed the intervals from
the light curves with flares, defined as light curve bins where there
is a $>5\%$ deviation between fits with flares and that of the underlying power
law, we find that upon refitting $L_{X,200\textrm{s}}$ and
$\alpha_{X,avg,>200\textrm{s}}$, the correlation gets even tighter.
Therefore, from 
here on out, we use the flare-removed average decay fits.

\subsection{X-ray Plateaus}

\begin{figure}
\includegraphics[width=0.5\textwidth,clip=true,trim=0 50 0 50]{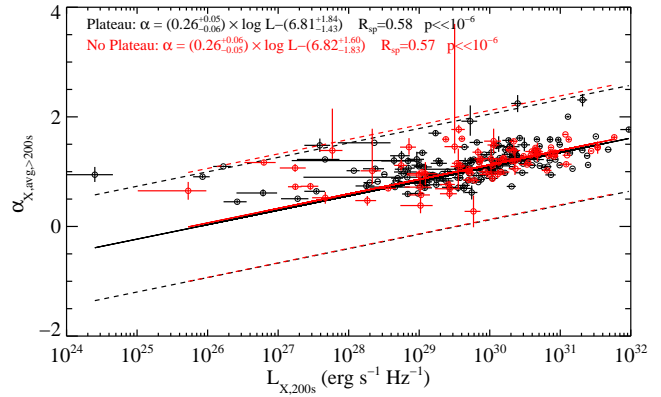}
\caption{Average decay - luminosity correlation using the sample
  including steep decay and flare corrections, those only of long duration, and
  split into those light curves that contain a plateau in
  their best fit (power-law) models and those that do not.  
  The correlation strengths and regression lines are
  consistent between the subsamples, suggesting that 
  plateaus do not influence the correlation significantly.  Therefore,
  light curves both with and without plateaus remain in our sample.
  The solid line indicates the best fit regression, and the dashed lines indicates
  the $2 \sigma$ deviation. 
  \label{fig:decaylum_plateau}}
\end{figure}

The complex light curve morphology of X-ray afterglows (Figure
\ref{fig:canon}-\ref{fig:steep_contam}) likely influences the average decay rates.  For
example, a light curve with an extremely long plateau (e.g. GRB
060729, \citealt{grupe07,grupe10}) may have a shallower average decay,
whereas GRBs without plateaus would be steeper.  
\cite{dainotti10,dainotti13} have shown that there is a relationship
between the time and flux of the end of the X-ray plateau, which could
be another manifestation of the behavior we have found in this study.
We test the effect of X-ray plateaus on the average decay-luminosity
correlation by separating our sample into those light curves that show
plateau behavior (defined as containing segment II in criteria set by
\citealt{racusin09}), and those that do not show clear plateaus.
Figure \ref{fig:decaylum_plateau} demonstrates that the average
decay-luminosity correlation is significant in both the sample with
and without plateaus.  This suggests that the presence of a plateau is
not necessarily solely responsible for regulating the average
afterglow decay.

\subsection{Final Correlation}

\begin{figure}
\includegraphics[width=0.5\textwidth,clip=true,trim=0 50 0 50]{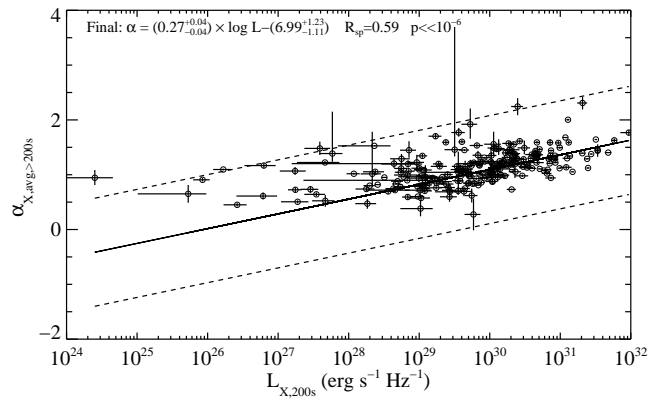}
\caption{Final average decay - luminosity correlation using the sample
  that includes steep decay and flares corrections and those only of
  long duration, with corrections and sub-sample optimization
  described in \S\ref{sec:results}.  The solid line indicates the best
  fit regression, and the dashed lines indicates the $2 \sigma$
  deviation.\label{fig:decaylum_final}}
\end{figure}

After correcting for steep decay contamination, filtering on long
duration GRBs, and testing for contamination by other light curve
morphological features, the final correlation is presented in Figure
\ref{fig:decaylum_final}.  Much of the scatter in the correlation has
been reduced.

However, a small population of shallow average decay, low luminosity
GRBs all occupy a parameter space to the left of
the best fit regression line. 
These include a number of ``low luminosity'' GRBs
\citep{sazonov04,soderberg06,liang07} at very low
redshifts ($z<0.3$).  The prompt and afterglow emission from these
objects tends to be much smoother than their high redshift/high
luminosity cousins.  Perhaps this represents some sort of asymptotic
value of the average slope, with very different X-ray afterglow
properties from typical canonical afterglows, perhaps indicative of
different jet dynamics or geometry.  This small sample of low luminosity GRBs
  have large errors in $L_{X,200\textrm{s}}$, and do not significantly
  drive the slope or significance of the correlation.

\subsection{Individual Segments\label{sec:indiv}}
\begin{figure}
\includegraphics[width=0.5\textwidth,clip=true,trim=0 50 0 50]{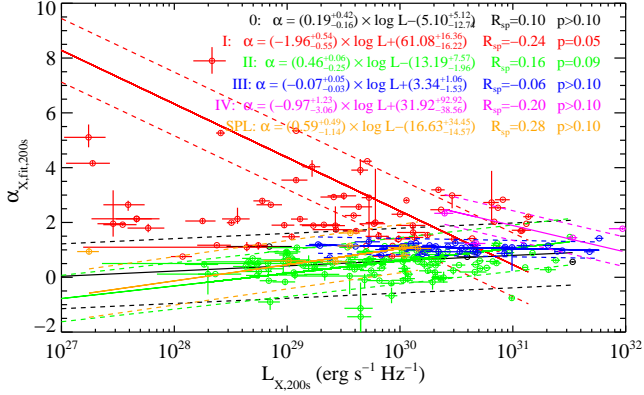}
\caption{Luminosity - Individual segment decay correlation test, where
  the temporal slopes and luminosities are interpolated from the
  best fit (power-law) models (for whichever segment dominates
  at $t_{200\textrm{s}}$), rather than the average decay.
  The roman numerals refer to the segments described in Figure
  \ref{fig:canon} of the best fit (power-law) models, and SPL refers to
    afterglows consisting of only a single power law.  There are no
    significant correlations present, demonstrating that no individual
    segment is driving the average decay-luminosity correlation, but
    rather an average energy output of the afterglow.  The solid
    line indicates the best fit regression, and the dashed lines
    indicates the $2 \sigma$ deviation. \label{fig:decaylum_fit}}
\end{figure}

The average decay rate of an X-ray light curve 
characterizes the afterglow differently than the slope of each
individual light curve segment.  The individual segments are
more directly influenced 
by details of the GRB environment, microphysics, and evolution of the
forward shock synchrotron spectrum \citep{sari98,granot02}, as
frequently demonstrated via the closure relations
\citep{panaitescu06,zhang06,racusin09}.  To demonstrate that the
  average decay rate correlation is significantly stronger than any
  correlation between the slopes of any individual segments, we
  conduct three tests: 1)  we measure the correlation between $L_{x,200\textrm{s}}$ and
the slope of the light curves at $t_{200s}$ from the best fit
  (power-law) models ($\alpha_{X,fit,200\textrm{s}}$) (Figure
\ref{fig:decaylum_fit}), where each
segment refers to those described in Figure 
\ref{fig:canon} ; and 2) we measure the correlation between
  $L_{x,200\textrm{s}}$ and the slope of the plateau (segment II; if present) even
  if not observed at $t_{200\textrm{s}}$ ($\alpha_{X,II}$); 3) we
  measure the correlation between 
  $L_{x,200\textrm{s}}$ and the slope of the normal decay phase
  (segment III; if present) even
  if not observed at $t_{200\textrm{s}}$ ($\alpha_{X,III}$).  None of the correlations 
between $L_{x,200\textrm{s}}$ and $\alpha_{X,fit,200\textrm{s}}$
  or $\alpha_{X,II}$ or $\alpha_{X,III}$
are significant for any of the segments of the canonical light curve,
implying the importance of the average decay 
measure.  All correlation values are given in Table \ref{table:corr}.

\section{Discussion} \label{sec:disc}

The following section demonstrates the robustness of the average decay
- luminosity correlation, its uniqueness among other afterglow
correlations, and excludes some observational biases.


\subsection{Observational Biases}

Hypothetically, the longer \swift observes a GRB, the more significant
the contribution of the normal forward shock decay and post jet-break
decay are to the average decay, dragging a fit to the average decay
index slightly steeper.  For any given canonical light curve
(i.e. Figure \ref{fig:canon}), a jet break within an afterglow is also
most likely to be detected if the GRB starts out brighter, allowing
the flux post-jet break to be higher relative to the XRT threshold.
We test the hypothesis that these effects could be driving the average
decay - luminosity correlation. 

Additionally, the observing policy of the \swift mission has evolved
over the last decade further complicating these trends, as \swift has
transitioned from a primarily GRB mission to an oversubscribed
general purpose monitoring and rapid-response observatory, which has
reduced the typical time spent on GRB follow-up.  In the last few
years, GRBs are generally dropped from the observing plan after the
first day of observations, unless they are especially remarkable and
interesting for coordinated follow-up observations.

\begin{figure}
\includegraphics[width=0.5\textwidth,clip=true,trim=0 50 0 50]{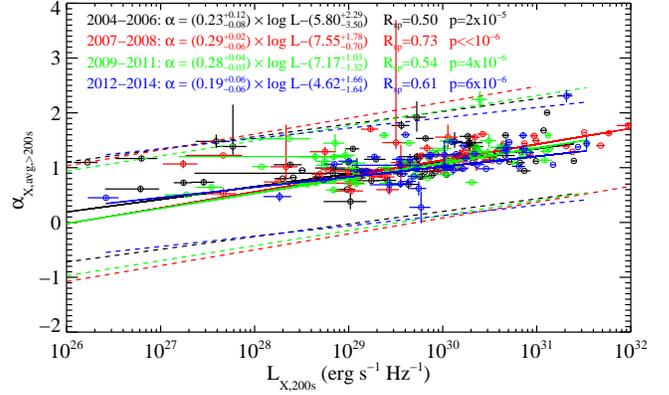}
\caption{The average decay-luminosity correlation evaluated for our
  sample split by when the GRBs were detected within the \swift
  mission.  The correlation strength is similar throughout the \swift
  mission,  demonstrating that the shifting
  mission policies of reducing GRB observation duration has minmal
  impact on the correlation strength.   The solid line indicates the best
  fit regression, and the dashed lines indicates the $2 \sigma$ deviation.
\label{fig:years_corr}}
\end{figure}

We evaluate the idea that changes in the duration of observations
influenced the average decay - luminosity correlation.  We look at the
trend of reducing GRB afterglow monitoring duration 
which has decreased with the years of operation of the \swift mission
at a rate of $\sim5-10\%$ per year.  There is a slight trend of
steeper $\alpha_{X,avg>200s}$ with GRBs that were observed for a
longer time post trigger, but that may be a real effect as the faster
fading GRBs become undetectable sooner after their triggers so are
therefore observed for a shorter time.
To see the impact of this trend on the average decay-luminosity
correlation, we split our sample into roughly equal sized subdivisions
by year of their detection (Figure \ref{fig:years_corr}).  The
correlation strength remains significant and similar for each
sub-sample.  Therefore, the reduction of observation duration has
minimal influence on the correlation.


Another potential observational source of the average
decay-luminosity correlation is that redshift effects are driving the
correlation via GRBs at higher redshift are being observed to have lower
initial flux values, and therefore they are less likely to be observed
at late times.  The longer an X-ray afterglow light curve is observed,
the more the average decay is affected by the contributions from the
steeper components.  In this scenario, the brightest GRBs would be
more likely to have observed jet breaks and therefore steeper decay
rates, where as the initially fainter GRBs would be less likely to be
observed at late times and have shallower decays.  We test this
hypothesis by
conducting Monte Carlo simulations in which we randomize both the 
redshift and flux normalization of each burst in our sample, then scale each light
curve accordingly (new distance, k-correction, $t_{200s}$).  We
  then fit the
average decay rate to the light curve points that occur after the new
$t_{200s}$ or end of steep decay, and the points that fall above a
fixed conservative XRT threshold of $1\times10^{-3}\ \textrm{counts}\
\textrm{s}^{-1}$.  We also extract a new $L_{X,200\textrm{s}}$ from the
appropriate quantities.  The entire sample is randomized $10^4$ times,
and for each set of randomizations, we extract $L_{X,200\textrm{s}}$ and
$\alpha_{X,avg,>200\textrm{s}}$ and measure the correlation strength
in the same manner as in \S\ref{sec:results}.  Of the $10^4$
simulations, none of the $R_{sp}$ coefficients are anywhere near the
value from our measured correlations. 


\subsection{Robustness of Average Decay-Luminosity Correlation
  Compared to Other GRB Correlations}
There are other correlations in the literature between various
afterglow quantities, which we recreate with our sample and
measurements for comparison, validation, and discussion of whether or
not these correlations are measurement of the same energy dissipation
mechanism. 

\begin{figure}
\includegraphics[width=0.5\textwidth,clip=true,trim=0 50 0 50]{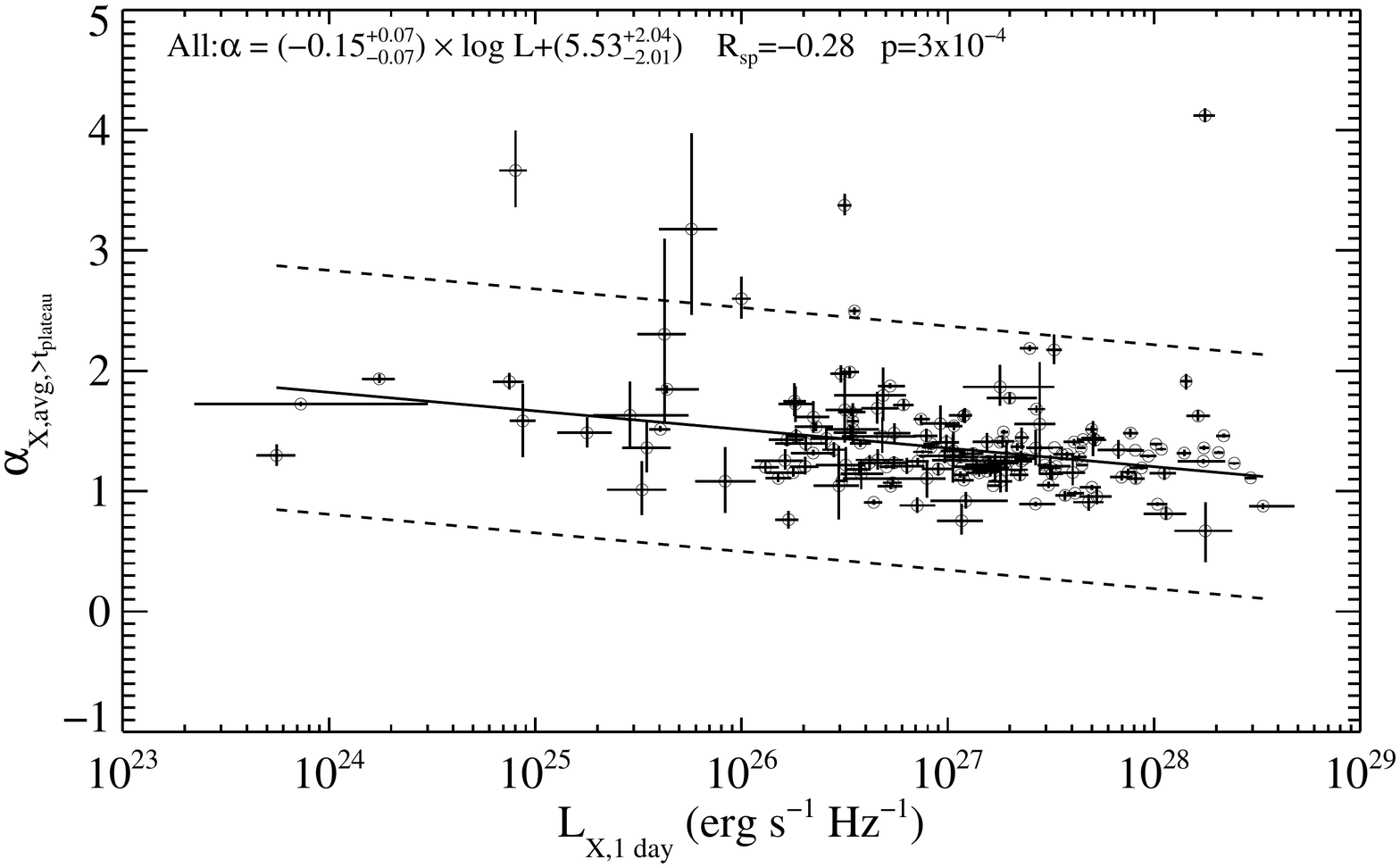}
\caption{Recreation of the luminosity at 1 day versus average decay
  after plateau correlation discussed in
  \cite{boer00,gendre05,gendre08,bardho15}.  We find a correlation in the
  opposite direction to that of these authors, which makes sense in
  the context of the correlation discussed in this paper, and
  demonstrates why measuring the correlation at early times is more
  robust.  The solid line indicates the best
  fit regression, and the dashed lines indicates the $2 \sigma$ deviation.
\label{fig:gendre}}
\end{figure}

\cite{dainotti10,dainotti13} describe a correlation between the time
and luminosity of X-ray afterglows at the end of their plateaus.  We
reproduce this measurement, confirming that the plateau
time-luminosity correlation is very significant in our sample, with
$R_{sp}=-0.71$ ($p=1.89\times10^{-26}$).  While Dainotti's correlation
has a similar or slightly stronger significance compared to the
average decay - luminosity, it is only applicable to the $\sim60\%$ of
GRBs with a plateau segment, therefore the average decay - luminosity
correlation is a more general description of afterglow behavior.

\begin{figure}
\centering
\includegraphics[width=0.48\textwidth,clip=true,trim=0 20 20 60]{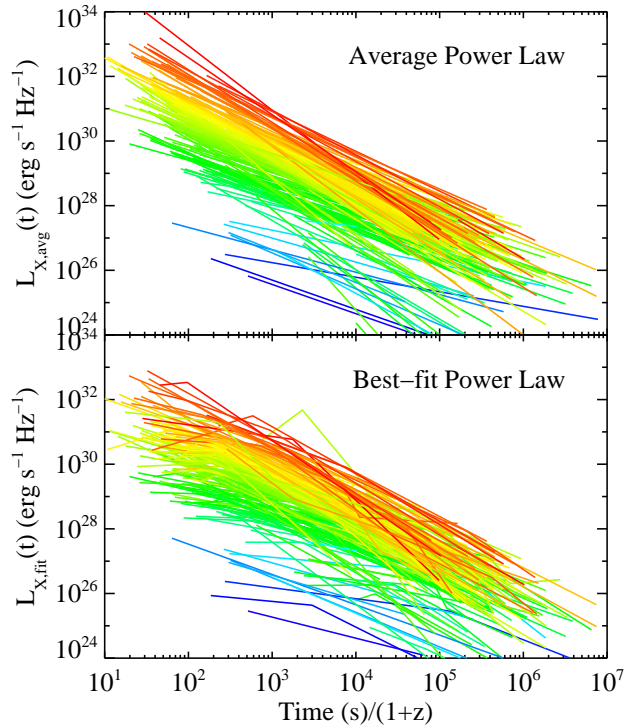}
\caption{The average power-law models to each long GRB afterglow (top panel) and
  best fit (power-law) models (bottom panel) colored by $L_{X,avg,200s}$ demonstrates
  the average decay - luminosity correlation qualitatively.  The
  color distinction is very clear at early times, and gets more mixed
  together at later times when the faster decay afterglows are no
  longer the brightest.  \label{fig:parrot}}
\end{figure}

\begin{figure}
\includegraphics[width=0.5\textwidth,clip=true,trim=0 50 0 50]{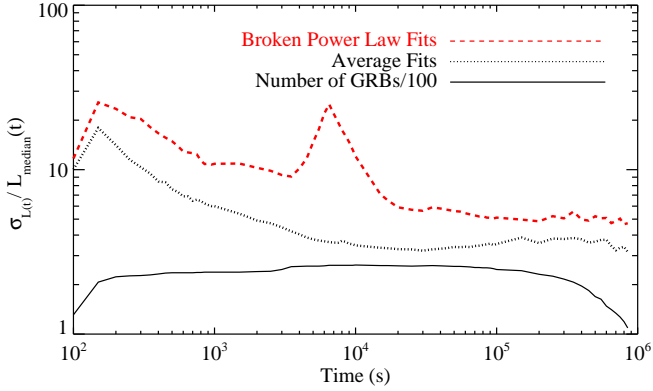}
\caption{The time averaged spread of the light curve fits (Figure
  \ref{fig:parrot}), measured as the standard deviation
  ($\sigma_{L(t)}$) of the luminosity distribution divided by the
  median luminosity as a function of time ($L_{median}(t)$).  The
  solid black curve indicates the scaled number of GRBs observed at
  those timescales used in the spread measurement, which drops off
  $<200\ \textrm{s}$ and $>3\times10^5\ \textrm{s}$ timescales, making
  them noisy.  The spread measurements are interpolated from the
  average (black dotted) and best fit (power-law) models (red
  dashed).  This parameterization demonstrates that the distribution
  of luminosities narrows and then widens with time, consistent with
  the picture that the fastest fading afterglows begin luminous and
  around $\sim10^4$ s they converge, and becoming the least luminous.
  This behavior is most obvious in the average power-law models, where as the
  best fit (power-law) models are muddied by the various durations of
  plateaus, causing the spike around $7\times10^4\
  \textrm{s}$.\label{fig:lumspread}}
\end{figure}

\cite{boer00,gendre05,gendre08,bardho15} describe early measurements of
BeppoSAX and the first years of XRT afterglow observations, and an
apparent clustering in luminosity space for which subsets had on
average different decay rates, suggesting that the more luminous GRBs
decay faster than the less luminous ones.  However, this correlation
became non-significant as a larger sample was accumulated
\citep{gendre08,bardho15}.  When we reconstruct their measurement, the decay
rate after the end of the plateau versus the luminosity at 1 day, with
our larger sample (Figure \ref{fig:gendre}), we find that there is a
correlation with $R_{sp}=-0.28$ ($p=3.04\times10^{-4}$).
However, we note that this correlation is actually going in the
opposite direction as previously claimed, and in the opposite
direction of the average decay-luminosity correlation (at $t_{200}$).
This can be easily reconciled by noting that as the more luminous
afterglows decay faster (early on), at some later point ($\sim1$ day),
the trend intersects, and afterwards on average, the faster decaying
afterglows have become the fainter ones (Figure \ref{fig:parrot}).  We
quantify this by measuring the spread of the luminosity distribution
of light curves as a function of time for both the best fit
  (power-law) models and average power-law models to our light curves (Figure 
\ref{fig:lumspread}).  In both cases, though more clearly in the
average power-law models, the spread decreases with time initially until
$\sim10^4$ s and reverses the trend increasing afterwards.

\subsection{Origin of the Average Decay - Luminosity Correlation}
As described in \cite{2012MNRAS.426L..86O,oates15}, the physical origin of the
correlation is somewhat uncertain, though \cite{oates15} explored the
 expectations for the luminosity - average decay correlation from 
the standard afterglow forward shock model \citep{sari98,zhang07} by
performing Monte Carlo simulations drawing randomly from distributions
of microphysical parameters: the kinetic energy of the outflow
  ($E_K$), the fraction of energy in the electric ($\epsilon_e$) and
  magnetic fields ($\epsilon_B$), the environmental density ($n$),
  and the electron spectral index (p). \cite{oates15} used the
  theoretical relationships (the closure relations;
  \cite{zhang06,racusin09,gao13}) between these parameters to derive
  the amplitude and shape of afterglow spectra and light curves, which
  in turn provided values for $L_{X,avg,200s}$ and
  $\alpha_{X,avg,>200\textrm{s}}$ for a sample the same size as their
  dataset.  They repeated this exercise for $10^4$ iterations, and
  determined that despite the simulated GRBs having similar
  X-ray-optical behavior, the Spearman rank coefficients for
  the optical and X-ray luminosity-average decay correlations were
  inconsistent with the observed values at a significance of
  $\gtrsim 4\sigma$.  The implication is that the observed
  correlations are much stronger than what is expected from
  theoretical models.


The average decay-luminosity correlation could be the result of
viewing geometry.  As \cite{vaneerten10,vaneerten11,ryan15} have shown
with their relativistic hydrodynamic simulations, jets viewed further
away from the jet axis (but inside the jet cone) show later jet
breaks, and fainter overall afterglow emission due to the role of jet
spreading.  This would suggest that those afterglows that start out
bright and fade sooner might be viewed more on-axis than those that
start out fainter and fade slower.  We tested this by searching for a
correlation between our early time luminosity, average decay, and the
off-axis observer angle for the GRBs in common with the sample from
\cite{ryan15}.  Unfortunately, we found no significant correlation
with the fractional observer angle.  Note that these simulations do
not include a mechanism for X-ray plateaus, and only fit light curves
after this phase, for which good measurements are only available on a
small subset of the sample \citep{ryan15}.  This does not rule out jet
geometry as the cause of the correlation, but does suggest that it
might be only one component of the mechanism that affects the average
GRB afterglow decay rate.

Another possible origin for the average decay-luminosity correlation
could be related the relationship between the theoretical
Synchrotron cooling timescale and the inverse of the strength of the
magnetic field and the bulk Lorentz factor ($\Gamma$):
$t_{sync}(\nu) \sim (B^3\Gamma)^{-0.5}$ (\citealt{piran04}, eq. 19).
If $t_{sync}$ is a proxy for the average decay rate (i.e. shorter cooling
timescale causes steeper average decay), then it should
scale with either large magnetic fields or large $\Gamma$, or both.
The cooling timescale is another way to parameterize the cooling
frequency $\nu_c$, which should be low (perhaps below the X-ray band)
if $\Gamma$ is very high, which for a constant density ISM
environment would cause the X-ray flux to be steep.
\cite{liang10} and \cite{ghirlanda12} have shown that there is a correlation
between $\Gamma$ and $E_{\gamma,iso}$ or $L_{\gamma,iso}$, and
\cite{gehrels08} and \cite{nysewander09} have shown a correlation between
$E_{\gamma,iso}$ and the afterglow luminosity, therefore it follows
that the luminosity and average decay rate should be linked.

\begin{figure}
\includegraphics[width=0.5\textwidth,clip=true,trim=0 50 0 50]{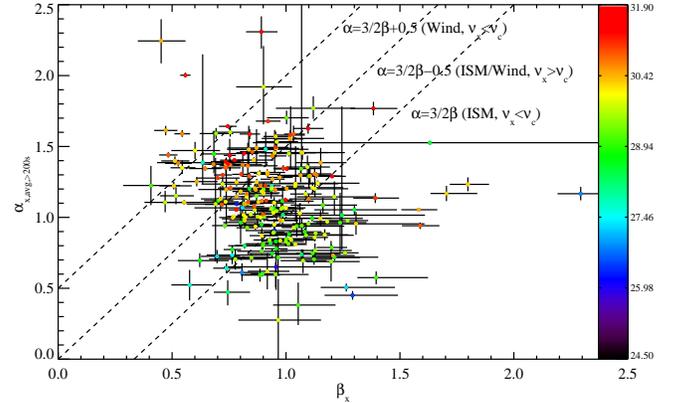}
\caption{The average temporal decay ($\alpha_{avg,x,>200s}$) and
  average spectral energy index ($\beta_x$) are compared with
  $log L_{x,200s}$ (color scale), demonstrating consistency and trends
  with the closure relations (dashed lines).  The high luminosity
  (redder) points are roughly consistent with Wind-like environments.   \label{fig:crs}}
\end{figure}

Yet another possible physical origin for the average decay-luminosity
correlation could be rooted in the circumburst environment.  We
look at $\alpha_{x,avg,>200s}$ and $L_{200s}$ in the context of the closure
relations \citep{panaitescu06,zhang06,racusin09} adding the spectral
energy index ($\beta_x$) for the standard forward shock (Figure
\ref{fig:crs}).  The highest luminosity GRBs tend toward the lines
demarcating the $R^{-2}$ wind environment with the X-ray band
($\nu_x$) below the synchrotron cooling frequency ($\nu_c$), and the
line showing either a constant density ISM or a Wind profile with the
X-ray band above the cooling frequency, and away from the ISM profile
with $\nu_x<\nu_c$.  The ambiguity in the $\nu_x>\nu_c$ cases
prohibits us from making a strong statement on the role of circumburst
environment, but it may be another possible contribution in that the
initially brightest GRB afterglows may be more likely to live in
Wind-like environments (see also \citealt{depasquale13}).

Based upon the optical correlation \citep{2012MNRAS.426L..86O,oates15}
alone, one could hypothesize that the most energetic GRBs show a
bright short-lived reverse shock that would cause the average
afterglow to be steeper.  However, the presence of the X-ray
correlation rules out this idea, as the reverse shock is expected to
produce strong emission in the optical band, but not the X-ray band
\citep{sari99}.

\section{Conclusions} \label{sec:conc}

In this paper, we show that there is a significant correlation between
early GRB X-ray afterglow luminosity and the average decay rate of
that afterglow, suggestive of some mechanism that moderates the energy
dissipation in GRB jets.  This mechanism could be related to the
interaction with the circumburst environment, physical parameters of
the jet, or geometrical effects.  The correlation is
undoubtably significant, and complementary to other afterglow
correlations demonstrated in the literature.  It is also only present
in the average decay and not in the individual segments of the
canonical afterglow light curve.  We argue that the
average decay - luminosity correlation is more robust than other
afterglow correlations as the decay rate is redshift independent, and it can be
applied to all long GRBs independent of their individual light curve
morphologies (e.g. with and without plateaus).

Although the precise origin of the correlation
remains unclear, GRB jet propagation model and numerical simulations
should strive to reproduce this observational feature.  The role of
geometry and viewing angle within the jet surely has a significant
impact on the observed emission, and disentangling that from intrinsic
physics remains difficult.

\cite{oates15} compared the UV/optical and X-ray average
decay - luminosity correlations, and explored many correlations with prompt
emission parameters ($T_{90}$, $E_{\gamma,iso}$), and determined that
a correlation as significant as the observed one could not be predicted with
the fireball model framework.

As redshift is a fundamental quantity in the luminosity measurement,
and the average decay is mostly redshift independent, the correlation
can be inverted to roughly estimate redshifts.  We attempt this
redshift estimation, but the scatter between measured redshift and
actual redshift deviate by $\Delta z\sim2\ (1 \sigma)$.  Therefore,
the intrinsic scatter in the average decay-luminosity correlation
prevents estimated redshifts with errors small enough to be of any use
in GRB physical parameter estimation or cosmology.  Perhaps more
detailed work on this topic could improve this estimation technique,
which could be explored in a future paper.

\section*{Acknowledgements}
The authors thank the anonymous referee for helpful comments, as well
as Brad Cenko and Raffaella Margutti for useful
discussions.  This work made use of data supplied by the UK {\it
  Swift} Science Data Centre at the University of Leicester. SRO
acknowledges the support of the Spanish Ministry, Project Number
AYA2012- 39727-C03-01.  MDP acknowledges the support
of the UK Space Agency.  DK acknowledges the support of the NASA
Postdoctoral Program.

\LTcapwidth=\textwidth
\begin{longtable*}{lcccccccc}
\toprule
GRB & z & z reference & $T_{90}$ &$\alpha_{X,avg,>200s}$ & $L_{X,200s}$ & Steep Decay
& Plateau & Flares \\
& & & (s) & & (erg cm$^{-2}$ Hz$^{-1}$) & Contaminated & &  \\
\midrule
\endhead
\toprule \multicolumn{9}{c}{{Continued on next page}} \\ \bottomrule
\endfoot
\endlastfoot
\input{grb_table}
\toprule
\caption{The properties of the GRBs in our sample including our
  measurements of the correlation parameters and where they are
  divided in our various sample subsets.  Steep decay contamination
  refers to those light curves where $t_{200s}$ occurs prior to the
  end of the steep decay phase, so we extrapolate the next segment
  back to $t_{200s}$.  Plateau/Flares refers to whether or not they
  are present in the X-ray light curve.
  \label{table:everything}}
\end{longtable*}

\bibliographystyle{apj}
\bibliography{bib,grb_z}

\end{document}

%% file: results_table.tex
Initial & $log\ L_{X,200s}$ & $\alpha_{X,avg,>200s}$ & 0.44 & $\ll 10^{-6}$ & 0.49 & $\ll 10^{-6}$ & $0.29_{-0.06}^{+0.03}$ & $-7.47_{-0.98}^{+1.80}$ & 246 \\[2pt]
\hline
Steep Decay Contam & $log\ L_{X,200s}$ & $\alpha_{X,avg,>200s}$ & -0.44 & $0.04$ & -0.10 & $> 0.10$ & $-0.83_{-0.34}^{+0.48}$ & $25.72_{-14.49}^{+9.69}$ & 23 \\[2pt]
Not Contam & $log\ L_{X,200s}$ & $\alpha_{X,avg,>200s}$ & 0.56 & $\ll 10^{-6}$ & 0.57 & $\ll 10^{-6}$ & $0.29_{-0.30}^{+0.04}$ & $-7.52_{-1.13}^{+8.97}$ & 223 \\[2pt]
\hline
Short & $log\ L_{X,200s}$ & $\alpha_{X,avg,>200s}$ & -0.07 & $> 0.10$ & 0.11 & $> 0.10$ & $0.16_{-0.39}^{+0.10}$ & $-3.40_{-2.77}^{+11.16}$ & 9 \\[2pt]
Long & $log\ L_{X,200s}$ & $\alpha_{X,avg,>200s}$ & 0.59 & $\ll 10^{-6}$ & 0.59 & $\ll 10^{-6}$ & $0.27_{-0.04}^{+0.04}$ & $-6.99_{-1.10}^{+1.23}$ & 237 \\[2pt]
\hline
Flares & $log\ L_{X,200s}$ & $\alpha_{X,avg,>200s}$ & 0.58 & $\ll 10^{-6}$ & 0.56 & $\ll 10^{-6}$ & $0.30_{-0.06}^{+0.07}$ & $-7.91_{-2.17}^{+1.84}$ & 134 \\[2pt]
No Flares & $log\ L_{X,200s}$ & $\alpha_{X,avg,>200s}$ & 0.59 & $\ll 10^{-6}$ & 0.64 & $\ll 10^{-6}$ & $0.28_{-0.04}^{+0.03}$ & $-7.27_{-1.01}^{+1.28}$ & 103 \\[2pt]
\hline
Plateau & $log\ L_{X,200s}$ & $\alpha_{X,avg,>200s}$ & 0.58 & $\ll 10^{-6}$ & 0.55 & $\ll 10^{-6}$ & $0.26_{-0.06}^{+0.05}$ & $-6.81_{-1.43}^{+1.84}$ & 156 \\[2pt]
No Plateau & $log\ L_{X,200s}$ & $\alpha_{X,avg,>200s}$ & 0.57 & $\ll 10^{-6}$ & 0.61 & $\ll 10^{-6}$ & $0.26_{-0.05}^{+0.06}$ & $-6.82_{-1.83}^{+1.60}$ & 81 \\[2pt]
\hline
{\bf Final} & $\mathbf{ log\ L_{X,200s}}$ & \boldmath$\alpha_{X,avg,>200s}$ & {\bf 0.59} & $\mathbf{ \ll 10^{-6}}$ & {\bf 0.59} & $\mathbf{ \ll 10^{-6}}$ & $\mathbf{ 0.27_{-0.04}^{+0.04}}$ & $\mathbf{ -6.99_{-1.11}^{+1.23}}$ & {\bf 237} \\[2pt]
\hline
\hline
0 & $log\ L_{X,200s}$ & $\alpha_{X,fit,200s}$ & 0.10 & $> 0.10$ & -0.58 & $> 0.10$ & $0.19_{-0.16}^{+0.42}$ & $-5.10_{-12.74}^{+5.12}$ & 5 \\[2pt]
I & $log\ L_{X,200s}$ & $\alpha_{X,fit,200s}$ & -0.24 & $0.05$ & -0.08 & $> 0.10$ & $-1.96_{-0.55}^{+0.54}$ & $61.08_{-16.22}^{+16.36}$ & 70 \\[2pt]
II & $log\ L_{X,200s}$ & $\alpha_{X,fit,200s}$ & 0.16 & $0.09$ & 0.15 & $> 0.10$ & $0.46_{-0.25}^{+0.06}$ & $-13.19_{-1.96}^{+7.57}$ & 113 \\[2pt]
III & $log\ L_{X,200s}$ & $\alpha_{X,fit,200s}$ & -0.06 & $> 0.10$ & -0.07 & $> 0.10$ & $-0.07_{-0.03}^{+0.05}$ & $3.34_{-1.53}^{+1.06}$ & 36 \\[2pt]
IV & $log\ L_{X,200s}$ & $\alpha_{X,fit,200s}$ & -0.20 & $> 0.10$ & -0.95 & $0.05$ & $-0.97_{-3.06}^{+1.23}$ & $31.92_{-38.56}^{+92.92}$ & 4 \\[2pt]
SPL & $log\ L_{X,200s}$ & $\alpha_{X,fit,200s}$ & 0.28 & $> 0.10$ & 0.30 & $> 0.10$ & $0.59_{-1.14}^{+0.49}$ & $-16.63_{-14.57}^{+34.45}$ & 9 \\[2pt]
\hline
Only II & $log\ L_{X,200s}$ & $\alpha_{X,fit,II} $ & 0.23 & $1\times 10^{-3}$ & 0.28 & $9\times 10^{-5}$ & $0.37_{-0.21}^{+0.05}$ & $-10.26_{-1.63}^{+6.28}$ & 194 \\[2pt]
Only III & $log\ L_{X,200s}$ & $\alpha_{X,fit,III}$ & 0.04 & $> 0.10$ & 0.03 & $> 0.10$ & $0.04_{-0.07}^{+0.07}$ & $-0.06_{-2.10}^{+2.22}$ & 156 \\[2pt]
\hline
2004-2006 & $log\ L_{X,200s}$ & $\alpha_{X,avg,>200s}$ & 0.50 & $2\times 10^{-5}$ & 0.50 & $1\times 10^{-5}$ & $0.23_{-0.08}^{+0.12}$ & $-5.80_{-3.50}^{+2.29}$ & 67 \\[2pt]
2007-2008 & $log\ L_{X,200s}$ & $\alpha_{X,avg,>200s}$ & 0.73 & $\ll 10^{-6}$ & 0.72 & $\ll 10^{-6}$ & $0.29_{-0.06}^{+0.02}$ & $-7.55_{-0.70}^{+1.78}$ & 60 \\[2pt]
2009-2011 & $log\ L_{X,200s}$ & $\alpha_{X,avg,>200s}$ & 0.54 & $4\times 10^{-6}$ & 0.53 & $7\times 10^{-6}$ & $0.28_{-0.03}^{+0.04}$ & $-7.17_{-1.32}^{+1.03}$ & 64 \\[2pt]
2012-2014 & $log\ L_{X,200s}$ & $\alpha_{X,avg,>200s}$ & 0.61 & $6\times 10^{-6}$ & 0.61 & $7\times 10^{-6}$ & $0.19_{-0.06}^{+0.06}$ & $-4.62_{-1.64}^{+1.66}$ & 46 \\[2pt]
\hline
All & $log\ L_{X,1 day}$ & $\alpha_{X,avg,>t_{plat}}$ & -0.28 & $3\times 10^{-4}$ & -0.28 & $3\times 10^{-4}$ & $-0.15_{-0.07}^{+0.07}$ & $5.53_{-2.01}^{+2.04}$ & 163 \\[2pt]
\hline

%% file: grb_table.tex
GRB050126  & 1.29 & \cite{2005ApJ...629..328B} & 48.0 & $0.96 \pm 0.21$ & $(1.22 \pm 0.55) \times 10^{29}$ & n & n & n \\                                                       
GRB050315  & 1.95 & \cite{2005ApJ...634..501B} & 95.4 & $0.60 \pm 0.02$ & $(9.59 \pm 2.19) \times 10^{28}$ & n & y & n \\                                                       
GRB050319  & 3.24 & \cite{2009ApJS..185..526F} & 151.6 & $0.84 \pm 0.02$ & $(5.72 \pm 1.01) \times 10^{29}$ & n & y & n \\                                                      
GRB050401  & 2.90 & \cite{2009ApJS..185..526F} & 33.3 & $1.09 \pm 0.02$ & $(6.63 \pm 0.87) \times 10^{30}$ & n & y & n \\                                                       
GRB050416A & 0.65 & \cite{2007ApJ...661..982S} & 6.7 & $0.83 \pm 0.01$ & $(4.69 \pm 0.56) \times 10^{28}$ & n & y & n \\                                                        
GRB050502B & 5.20 & \cite{2011AA...526A.154A} & 17.7 & $0.87 \pm 0.05$ & $(1.10 \pm 0.12) \times 10^{30}$ & n & n & y \\                                                        
GRB050525A & 0.61 & \cite{GCN3483} & 8.8 & $1.53 \pm 0.04$ & $(6.58 \pm 1.40) \times 10^{29}$ & n & y & n \\                                                                    
GRB050724  & 0.26 & \cite{GCN3700} & 98.7 & $0.72 \pm 0.05$ & $(1.75 \pm 0.39) \times 10^{27}$ & n & n & y \\                                                                   
GRB050730  & 3.97 & \cite{2009ApJS..185..526F} & 154.6 & $2.00 \pm 0.02$ & $(1.29 \pm 0.07) \times 10^{31}$ & n & y & y \\                                                      
GRB050801  & 1.38 & \cite{2009MNRAS.395..490O} & 19.6 & $1.15 \pm 0.04$ & $(1.95 \pm 0.52) \times 10^{29}$ & n & y & n \\                                                       
GRB050802  & 1.71 & \cite{2009ApJS..185..526F} & 27.5 & $1.18 \pm 0.01$ & $(1.41 \pm 0.11) \times 10^{30}$ & n & y & y \\                                                       
GRB050814  & 5.30 & \cite{2006AIPC..836..552J} & 142.9 & $0.94 \pm 0.02$ & $(8.42^{+1.06}_{-1.02}) \times 10^{29}$ & n & y & y \\                                               
GRB050819  & 2.50 & \cite{2012ApJ...758...46K} & 37.7 & $0.38 \pm 0.16$ & $(1.04 \pm 0.51) \times 10^{29}$ & n & n & y \\                                                       
GRB050820A & 2.61 & \cite{2009ApJS..185..526F} & 240.8 & $1.22 \pm 0.01$ & $(4.60 \pm 1.23) \times 10^{30}$ & n & y & y \\                                                      
GRB050822  & 1.43 & \cite{2012ApJ...756..187H} & 104.3 & $0.81 \pm 0.02$ & $(1.30 \pm 0.22) \times 10^{29}$ & n & y & y \\                                                      
GRB050826  & 0.30 & \cite{2007ApJ...661L.127M} & 29.6 & $1.05 \pm 0.04$ & $(2.38 \pm 0.79) \times 10^{28}$ & n & y & n \\                                                       
GRB050904  & 6.29 & \cite{2006Natur.440..184K} & 181.6 & $1.29 \pm 0.03$ & $(6.85 \pm 0.53) \times 10^{30}$ & n & n & y \\                                                      
GRB050908  & 3.34 & \cite{2009ApJS..185..526F} & 18.3 & $1.23 \pm 0.05$ & $(1.52 \pm 0.38) \times 10^{30}$ & n & y & y \\                                                       
GRB050915A & 2.53 & \cite{2012ApJ...758...46K} & 53.4 & $1.20 \pm 0.07$ & $(5.32 \pm 1.86) \times 10^{29}$ & n & y & y \\                                                       
GRB050922C & 2.20 & \cite{2009ApJS..185..526F} & 4.6 & $1.30 \pm 0.02$ & $(3.82 \pm 0.37) \times 10^{30}$ & n & y & n \\                                                        
GRB051001  & 2.43 & \cite{2012ApJ...758...46K} & 190.3 & $0.77 \pm 0.06$ & $(2.59 \pm 0.62) \times 10^{29}$ & y & n & n \\                                                      
GRB051006  & 1.06 & \cite{2012ApJ...756..187H} & 35.4 & $1.77 \pm 0.08$ & $(3.64 \pm 0.79) \times 10^{29}$ & n & n & n \\                                                       
GRB051016B & 0.94 & \cite{GCN4186} & 4.0 & $0.80 \pm 0.02$ & $(1.99 \pm 0.37) \times 10^{28}$ & n & y & n \\                                                                    
GRB051109A & 2.35 & \cite{GCN4221} & 36.2 & $1.19 \pm 0.01$ & $(3.34 \pm 0.45) \times 10^{30}$ & n & y & n \\                                                                   
GRB051109B & 0.08 & \cite{GCN5387} & 15.7 & $0.91 \pm 0.03$ & $(8.47 \pm 2.24) \times 10^{25}$ & n & y & n \\                                                                   
GRB051117B & 0.48 & \cite{2012ApJ...756..187H} & 9.0 & $1.39^{+0.76}_{-0.15}$ & $(5.87^{+2.30}_{-2.02}) \times 10^{27}$ & n & n & n \\                                          
GRB051221A & 0.55 & \cite{GCN4384} & 1.4 & $0.94 \pm 0.02$ & $(6.61 \pm 2.07) \times 10^{27}$ & n & y & y \\                                                                    
GRB051227  & 0.80 & \cite{GCN4409} & 115.4 & $1.48 \pm 0.12$ & $(3.90 \pm 1.56) \times 10^{27}$ & y & y & y \\                                                                  
GRB060108  & 2.70 & \cite{GCN4539} & 14.2 & $0.76 \pm 0.03$ & $(1.74 \pm 0.36) \times 10^{29}$ & n & y & y \\                                                                   
GRB060111A & 2.32 & Perley Host\footnote{\url{http://www.astro.caltech.edu/grbhosts/redshifts.html}} & 13.2 & $0.84 \pm 0.03$ & $(6.06 \pm 1.09) \times 10^{29}$ & n & n & y \\ 
GRB060115  & 3.53 & \cite{2009ApJS..185..526F} & 139.1 & $0.86 \pm 0.03$ & $(7.59 \pm 0.96) \times 10^{29}$ & n & y & y \\                                                      
GRB060124  & 2.30 & \cite{2009ApJS..185..526F} & 13.4 & $1.12 \pm 0.01$ & $(9.79 \pm 0.63) \times 10^{30}$ & n & y & y \\                                                       
GRB060202  & 0.78 & \cite{2013ApJ...778..128P} & 192.9 & $0.94 \pm 0.02$ & $(7.38 \pm 1.24) \times 10^{30}$ & n & y & y \\                                                      
GRB060206  & 4.05 & \cite{2009ApJS..185..526F} & 7.6 & $1.24 \pm 0.02$ & $(8.49 \pm 2.82) \times 10^{29}$ & y & y & n \\                                                        
GRB060210  & 3.91 & \cite{2009ApJS..185..526F} & 288.0 & $1.12 \pm 0.02$ & $(2.48 \pm 0.13) \times 10^{31}$ & n & y & y \\                                                      
GRB060218  & 0.03 & \cite{GCN4792} & 80.0 & $1.17 \pm 0.05$ & $(6.27 \pm 2.95) \times 10^{26}$ & y & y & y \\                                                                   
GRB060223A & 4.41 & \cite{GCN4815} & 11.3 & $1.92^{+0.29}_{-0.26}$ & $(5.30 \pm 1.34) \times 10^{29}$ & n & y & y \\                                                            
GRB060319  & 1.17 & \cite{2013ApJ...778..128P} & 10.3 & $0.99 \pm 0.02$ & $(6.83 \pm 2.44) \times 10^{28}$ & n & y & y \\                                                       
GRB060418  & 1.49 & \cite{GCN4969} & 109.1 & $1.40 \pm 0.02$ & $(1.92 \pm 0.25) \times 10^{30}$ & n & y & y \\                                                                  
GRB060502A & 1.51 & \cite{2009ApJS..185..526F} & 508.1 & $0.89 \pm 0.02$ & $(3.49 \pm 0.56) \times 10^{29}$ & n & y & n \\                                                      
GRB060510A & 1.20 & \cite{2012MNRAS.426L..86O} & 20.1 & $1.01 \pm 0.01$ & $(1.19 \pm 0.11) \times 10^{30}$ & n & y & n \\                                                       
GRB060510B & 4.94 & \cite{GCN5104} & 262.9 & $0.69 \pm 0.09$ & $(4.31 \pm 1.35) \times 10^{29}$ & n & y & y \\                                                                  
GRB060512  & 2.10 & \cite{2009ApJS..185..526F} & 8.4 & $1.17 \pm 0.05$ & $(1.35 \pm 0.26) \times 10^{30}$ & n & n & y \\                                                        
GRB060522  & 5.11 & \cite{GCN5155} & 69.1 & $1.39 \pm 0.10$ & $(2.20 \pm 0.50) \times 10^{30}$ & n & y & y \\                                                                   
GRB060526  & 3.22 & \cite{2009ApJS..185..526F} & 298.0 & $1.00 \pm 0.03$ & $(8.61 \pm 1.68) \times 10^{29}$ & n & y & y \\                                                      
GRB060604  & 2.14 & \cite{2012ApJ...758...46K} & 96.0 & $0.90 \pm 0.02$ & $(6.77 \pm 6.54) \times 10^{28}$ & n & y & y \\                                                       
GRB060605  & 3.78 & \cite{2009AA...497..729F} & 580.9 & $1.57 \pm 0.04$ & $(1.85 \pm 0.34) \times 10^{30}$ & n & y & y \\                                                       
GRB060607A & 3.07 & \cite{2009ApJS..185..526F} & 103.0 & $1.59 \pm 0.03$ & $(4.66 \pm 0.41) \times 10^{30}$ & n & y & y \\                                                      
GRB060614  & 0.13 & \cite{2006Natur.444.1050D} & 109.1 & $1.09 \pm 0.02$ & $(1.67 \pm 0.47) \times 10^{26}$ & y & y & y \\                                                      
GRB060707  & 3.42 & \cite{2009ApJS..185..526F} & 66.6 & $0.93 \pm 0.02$ & $(6.46 \pm 1.50) \times 10^{29}$ & y & y & y \\                                                       
GRB060708  & 1.92 & \cite{2009ApJS..185..526F} & 10.0 & $1.16 \pm 0.03$ & $(5.50 \pm 0.65) \times 10^{29}$ & n & y & n \\                                                       
GRB060714  & 2.71 & \cite{2009ApJS..185..526F} & 116.1 & $1.03 \pm 0.02$ & $(1.10 \pm 0.14) \times 10^{30}$ & n & y & y \\                                                      
GRB060719  & 1.53 & \cite{2012ApJ...758...46K} & 66.9 & $0.98 \pm 0.02$ & $(3.89 \pm 0.61) \times 10^{29}$ & n & y & y \\                                                       
GRB060729  & 0.54 & \cite{2009ApJS..185..526F} & 113.0 & $0.82 \pm 0.01$ & $(2.58 \pm 0.23) \times 10^{28}$ & n & y & y \\                                                      
GRB060814  & 1.92 & \cite{2012ApJ...758...46K} & 145.1 & $1.07 \pm 0.01$ & $(1.00 \pm 0.16) \times 10^{30}$ & n & y & y \\                                                      
GRB060904B & 0.70 & \cite{2009ApJS..185..526F} & 190.0 & $1.34 \pm 0.04$ & $(1.11 \pm 0.19) \times 10^{29}$ & n & y & y \\                                                      
GRB060906  & 3.69 & \cite{2009ApJS..185..526F} & 44.6 & $1.06 \pm 0.04$ & $(3.96 \pm 0.84) \times 10^{29}$ & n & y & y \\                                                       
GRB060908  & 1.88 & \cite{2009ApJS..185..526F} & 19.3 & $1.46 \pm 0.05$ & $(1.90 \pm 0.31) \times 10^{30}$ & n & y & n \\                                                       
GRB060912A & 0.94 & \cite{GCN5617} & 5.0 & $1.10 \pm 0.03$ & $(9.92 \pm 1.42) \times 10^{28}$ & n & y & n \\                                                                    
GRB060923A & 2.80 & \cite{2013ApJ...778..128P} & 58.5 & $1.10 \pm 0.02$ & $(4.47 \pm 1.26) \times 10^{29}$ & n & y & n \\                                                       
GRB060926  & 3.20 & \cite{2009ApJS..185..526F} & 8.8 & $1.46^{+0.16}_{-0.14}$ & $(1.94 \pm 0.70) \times 10^{30}$ & n & y & y \\                                                 
GRB060927  & 5.47 & \cite{2007ApJ...669....1R} & 22.4 & $1.48^{+0.17}_{-0.12}$ & $(1.34 \pm 0.29) \times 10^{30}$ & n & y & n \\                                                
GRB061006  & 0.44 & \cite{2007ApJ...664.1000B} & 129.8 & $0.73 \pm 0.06$ & $(2.87^{+0.84}_{-0.79}) \times 10^{27}$ & n & n & n \\                                               
GRB061007  & 1.26 & \cite{2009ApJS..185..526F} & 75.7 & $1.68 \pm 0.01$ & $(1.18 \pm 0.09) \times 10^{31}$ & n & n & n \\                                                       
GRB061021  & 0.35 & \cite{2009ApJS..185..526F} & 47.8 & $0.95 \pm 0.01$ & $(3.21 \pm 0.29) \times 10^{28}$ & n & y & n \\                                                       
GRB061110A & 0.76 & \cite{2009ApJS..185..526F} & 44.5 & $0.61 \pm 0.07$ & $(6.11 \pm 3.55) \times 10^{26}$ & y & y & y \\                                                       
GRB061121  & 1.31 & \cite{2009ApJS..185..526F} & 81.2 & $1.09 \pm 0.01$ & $(2.07 \pm 0.13) \times 10^{30}$ & n & y & y \\                                                       
GRB061222A & 2.09 & \cite{2009AJ....138.1690P} & 96.0 & $1.11 \pm 0.01$ & $(5.12 \pm 0.36) \times 10^{30}$ & n & y & y \\                                                       
GRB070103  & 2.62 & \cite{2012ApJ...758...46K} & 18.4 & $1.21 \pm 0.04$ & $(1.07 \pm 0.23) \times 10^{30}$ & n & y & y \\                                                       
GRB070110  & 2.35 & \cite{2009ApJS..185..526F} & 88.4 & $1.22 \pm 0.03$ & $(5.95 \pm 0.87) \times 10^{29}$ & n & y & y \\                                                       
GRB070129  & 2.34 & \cite{2012ApJ...758...46K} & 459.7 & $0.96 \pm 0.02$ & $(2.12 \pm 0.83) \times 10^{29}$ & y & y & y \\                                                      
GRB070208  & 1.16 & \cite{GCN6083} & 64.0 & $1.29 \pm 0.09$ & $(5.61 \pm 1.74) \times 10^{28}$ & n & y & y \\                                                                   
GRB070306  & 1.50 & \cite{2008ApJ...681..453J} & 209.2 & $0.93 \pm 0.02$ & $(1.66 \pm 0.30) \times 10^{29}$ & n & y & y \\                                                      
GRB070318  & 0.84 & \cite{GCN6217} & 130.4 & $1.05 \pm 0.04$ & $(5.02 \pm 0.43) \times 10^{29}$ & n & y & y \\                                                                  
GRB070411  & 2.95 & \cite{GCN6283} & 102.0 & $1.13 \pm 0.03$ & $(3.38 \pm 0.78) \times 10^{30}$ & n & y & n \\                                                                  
GRB070419A & 0.97 & \cite{GCN6322} & 160.0 & $1.46^{+2.24}_{-0.12}$ & $(3.18 \pm 0.90) \times 10^{29}$ & n & n & y \\                                                           
GRB070419B & 1.96 & \cite{2012ApJ...758...46K} & 238.0 & $1.44 \pm 0.02$ & $(7.39 \pm 0.58) \times 10^{30}$ & n & y & y \\                                                      
GRB070506  & 2.31 & \cite{GCN6379} & 6.0 & $0.60 \pm 0.09$ & $(2.69 \pm 0.73) \times 10^{29}$ & n & n & n \\                                                                    
GRB070508  & 3.00 & \cite{GCN6398} & 20.9 & $1.39 \pm 0.01$ & $(2.05 \pm 0.22) \times 10^{31}$ & n & y & n \\                                                                   
GRB070518  & 1.16 & Perley Host\footnotemark[\value{footnote}] & 5.5 & $0.74 \pm 0.04$ & $(1.78 \pm 0.54) \times 10^{28}$ & y & y & y \\                                        
GRB070521  & 1.35 & \cite{2013ApJ...778..128P} & 38.6 & $1.32 \pm 0.02$ & $(1.76 \pm 0.29) \times 10^{30}$ & n & y & y \\                                                       
GRB070529  & 2.50 & \cite{GCN6470} & 108.9 & $1.20 \pm 0.03$ & $(1.51 \pm 0.25) \times 10^{30}$ & n & y & n \\                                                                  
GRB070714A & 1.58 & Perley Host\footnotemark[\value{footnote}] & 3.0 & $0.87 \pm 0.06$ & $(2.15 \pm 0.74) \times 10^{29}$ & n & y & y \\                                        
GRB070714B & 0.92 & \cite{2009ApJ...690..231B} & 65.6 & $1.70 \pm 0.05$ & $(1.71 \pm 0.36) \times 10^{29}$ & y & y & n \\                                                       
GRB070721B & 3.63 & \cite{2009ApJS..185..526F} & 336.9 & $1.61 \pm 0.03$ & $(2.66 \pm 0.33) \times 10^{30}$ & n & y & y \\                                                      
GRB070724A & 0.46 & \cite{GCN6665} & 0.4 & $1.26 \pm 0.11$ & $(1.06 \pm 0.57) \times 10^{27}$ & y & y & y \\                                                                    
GRB070802  & 2.45 & \cite{2009ApJS..185..526F} & 15.8 & $0.80 \pm 0.04$ & $(1.15 \pm 0.25) \times 10^{29}$ & n & y & n \\                                                       
GRB070809  & 0.22 & \cite{GCN7889} & 1.3 & $0.52 \pm 0.05$ & $(4.23 \pm 1.14) \times 10^{26}$ & n & y & n \\                                                                    
GRB070810A & 2.17 & \cite{GCN6741} & 9.0 & $1.36 \pm 0.08$ & $(6.81 \pm 1.11) \times 10^{29}$ & n & y & n \\                                                                    
GRB071020  & 2.14 & \cite{2009ApJS..185..526F} & 4.3 & $1.22 \pm 0.03$ & $(1.83 \pm 0.27) \times 10^{30}$ & n & y & y \\                                                        
GRB071021  & 2.45 & \cite{2012ApJ...758...46K} & 228.7 & $0.83 \pm 0.03$ & $(2.41 \pm 0.65) \times 10^{29}$ & n & y & y \\                                                      
GRB071025  & 5.20 & \cite{2009ApJS..185..526F} & 241.3 & $1.63 \pm 0.04$ & $(5.83 \pm 0.52) \times 10^{31}$ & n & y & n \\                                                      
GRB071031  & 2.69 & \cite{2009ApJS..185..526F} & 180.6 & $1.13 \pm 0.09$ & $(1.31 \pm 0.37) \times 10^{30}$ & n & y & y \\                                                      
GRB071122  & 1.14 & \cite{GCN7124} & 71.4 & $3.10 \pm 0.33$ & $(1.26 \pm 0.29) \times 10^{29}$ & y & n & y \\                                                                   
GRB071227  & 0.38 & \cite{2009ApJ...690..231B} & 142.5 & $1.07 \pm 0.07$ & $(1.74^{+0.72}_{-0.68}) \times 10^{27}$ & n & n & n \\                                               
GRB080123  & 0.50 & \cite{2010ApJ...725.1202L} & 114.9 & $1.02^{+0.77}_{-0.24}$ & $(2.16 \pm 0.70) \times 10^{28}$ & n & n & n \\                                               
GRB080207  & 2.09 & \cite{2012ApJ...756..187H} & 292.5 & $1.77 \pm 0.05$ & $(9.38 \pm 2.27) \times 10^{31}$ & n & y & n \\                                                      
GRB080210  & 2.64 & \cite{2009ApJS..185..526F} & 42.3 & $1.30 \pm 0.05$ & $(2.20 \pm 0.27) \times 10^{30}$ & n & y & y \\                                                       
GRB080310  & 2.42 & \cite{2009ApJS..185..526F} & 363.2 & $1.07 \pm 0.03$ & $(4.08 \pm 0.92) \times 10^{29}$ & n & y & y \\                                                      
GRB080319B & 0.94 & \cite{2009ApJS..185..526F} & 124.9 & $1.64 \pm 0.01$ & $(3.20 \pm 0.13) \times 10^{31}$ & n & y & n \\                                                      
GRB080319C & 1.95 & \cite{2009ApJS..185..526F} & 29.6 & $1.38 \pm 0.03$ & $(4.43 \pm 0.53) \times 10^{30}$ & n & y & y \\                                                       
GRB080325  & 1.78 & \cite{2013ApJ...778..128P} & 166.7 & $0.99 \pm 0.06$ & $(1.44 \pm 0.45) \times 10^{30}$ & n & n & y \\                                                      
GRB080413A & 2.43 & \cite{2009ApJS..185..526F} & 46.4 & $1.56 \pm 0.23$ & $(1.14 \pm 0.31) \times 10^{30}$ & n & n & n \\                                                       
GRB080413B & 1.10 & \cite{2009ApJS..185..526F} & 8.0 & $1.02 \pm 0.01$ & $(7.12 \pm 0.51) \times 10^{29}$ & n & y & n \\                                                        
GRB080430  & 0.77 & \cite{GCN7654} & 13.9 & $0.78 \pm 0.01$ & $(5.39 \pm 0.52) \times 10^{28}$ & n & y & n \\                                                                   
GRB080517  & 0.09 & \cite{2015MNRAS.446.3911S} & 64.5 & $0.65 \pm 0.16$ & $(5.31 \pm 4.29) \times 10^{25}$ & y & n & y \\                                                       
GRB080520  & 1.54 & \cite{2009ApJS..185..526F} & 3.3 & $0.94 \pm 0.08$ & $(9.90^{+3.21}_{-3.19}) \times 10^{28}$ & n & n & n \\                                                 
GRB080603B & 2.69 & \cite{2009ApJS..185..526F} & 59.1 & $1.20^{+0.23}_{-0.26}$ & $(2.28 \pm 0.52) \times 10^{30}$ & n & n & n \\                                                
GRB080604  & 1.42 & \cite{GCN7818} & 77.6 & $0.52 \pm 0.11$ & $(4.66 \pm 1.65) \times 10^{27}$ & y & n & n \\                                                                   
GRB080605  & 1.64 & \cite{2009ApJS..185..526F} & 18.1 & $1.34 \pm 0.02$ & $(4.65 \pm 0.55) \times 10^{30}$ & n & y & n \\                                                       
GRB080607  & 3.04 & \cite{2009ApJS..185..526F} & 79.0 & $1.49 \pm 0.02$ & $(9.83 \pm 0.95) \times 10^{30}$ & n & y & y \\                                                       
GRB080707  & 1.23 & \cite{2009ApJS..185..526F} & 30.2 & $0.75 \pm 0.03$ & $(4.29 \pm 0.97) \times 10^{28}$ & n & y & n \\                                                       
GRB080721  & 2.59 & \cite{2009ApJS..185..526F} & 159.0 & $1.40 \pm 0.01$ & $(4.78 \pm 0.51) \times 10^{31}$ & n & y & n \\                                                      
GRB080804  & 2.20 & \cite{2009ApJS..185..526F} & 37.9 & $1.11 \pm 0.02$ & $(1.09 \pm 0.08) \times 10^{30}$ & n & n & n \\                                                       
GRB080805  & 1.50 & \cite{2009ApJS..185..526F} & 106.6 & $0.95 \pm 0.04$ & $(9.74^{+2.61}_{-2.59}) \times 10^{29}$ & n & n & y \\                                               
GRB080810  & 3.35 & \cite{2009ApJS..185..526F} & 107.7 & $1.58 \pm 0.03$ & $(9.37 \pm 0.91) \times 10^{30}$ & n & y & y \\                                                      
GRB080905A & 0.12 & \cite{2010MNRAS.408..383R} & 1.0 & $2.75 \pm 0.18$ & $(2.57 \pm 0.76) \times 10^{27}$ & n & y & n \\                                                        
GRB080905B & 2.37 & \cite{2009ApJS..185..526F} & 120.9 & $1.15 \pm 0.02$ & $(2.97 \pm 0.34) \times 10^{30}$ & n & y & n \\                                                      
GRB080906  & 2.10 & \cite{GCN8212} & 148.2 & $1.11 \pm 0.02$ & $(1.40^{+0.48}_{-0.43}) \times 10^{30}$ & n & y & y \\                                                           
GRB080913  & 6.70 & \cite{GCN8225} & 7.5 & $1.09^{+0.33}_{-0.41}$ & $(3.58^{+0.81}_{-0.79}) \times 10^{29}$ & n & n & y \\                                                      
GRB080916A & 0.69 & \cite{2009ApJS..185..526F} & 61.3 & $0.96 \pm 0.02$ & $(6.02 \pm 1.00) \times 10^{28}$ & n & y & y \\                                                       
GRB080928  & 1.69 & \cite{2009ApJS..185..526F} & 233.7 & $1.55 \pm 0.04$ & $(2.19 \pm 0.30) \times 10^{30}$ & n & y & y \\                                                      
GRB081007  & 0.53 & \cite{GCN8335} & 9.7 & $0.88 \pm 0.02$ & $(4.94 \pm 0.95) \times 10^{28}$ & n & y & y \\                                                                    
GRB081008  & 1.97 & \cite{GCN8346} & 187.8 & $1.25 \pm 0.02$ & $(1.58 \pm 0.18) \times 10^{30}$ & n & y & y \\                                                                  
GRB081028A & 3.04 & \cite{GCN8434} & 284.4 & $1.22 \pm 0.03$ & $(4.62 \pm 2.77) \times 10^{27}$ & y & y & y \\                                                                  
GRB081109  & 0.98 & \cite{2011AA...534A.108K} & 221.5 & $1.17 \pm 0.02$ & $(7.17 \pm 0.65) \times 10^{29}$ & n & y & n \\                                                       
GRB081118  & 2.58 & \cite{GCN8531} & 53.4 & $0.58 \pm 0.05$ & $(1.06 \pm 0.36) \times 10^{29}$ & n & n & n \\                                                                   
GRB081203A & 2.05 & \cite{GCN8601} & 223.0 & $1.43 \pm 0.02$ & $(4.44 \pm 0.37) \times 10^{30}$ & n & y & n \\                                                                  
GRB081221  & 2.26 & \cite{2012ApJ...749...68S} & 33.9 & $1.29 \pm 0.01$ & $(2.40 \pm 0.22) \times 10^{31}$ & n & y & y \\                                                       
GRB081222  & 2.77 & \cite{GCN8713} & 33.0 & $1.22 \pm 0.01$ & $(8.53 \pm 0.70) \times 10^{30}$ & n & y & n \\                                                                   
GRB081228  & 3.40 & \cite{GCN8752} & 3.0 & $0.72 \pm 0.14$ & $(1.03^{+0.43}_{-0.41}) \times 10^{29}$ & n & n & n \\                                                             
GRB081230  & 2.00 & \cite{2011AA...526A.153K} & 60.7 & $0.85 \pm 0.03$ & $(1.77 \pm 0.49) \times 10^{29}$ & n & y & y \\                                                        
GRB090102  & 1.55 & \cite{GCN8766} & 24.7 & $1.38 \pm 0.01$ & $(4.88 \pm 0.43) \times 10^{30}$ & n & y & n \\                                                                   
GRB090113  & 1.75 & \cite{2012ApJ...758...46K} & 9.1 & $1.29 \pm 0.06$ & $(2.44 \pm 0.54) \times 10^{30}$ & n & y & n \\                                                        
GRB090205  & 4.65 & \cite{GCN8892} & 8.8 & $1.10 \pm 0.04$ & $(1.17 \pm 0.18) \times 10^{30}$ & n & y & n \\                                                                    
GRB090401B & 3.10 & \cite{2012MNRAS.426L..86O} & 186.5 & $1.45 \pm 0.01$ & $(3.40 \pm 0.33) \times 10^{31}$ & n & y & n \\                                                      
GRB090407  & 1.45 & \cite{2012ApJ...758...46K} & 315.5 & $0.79 \pm 0.02$ & $(9.43 \pm 2.15) \times 10^{28}$ & n & y & y \\                                                      
GRB090417B & 0.34 & \cite{GCN9156} & 266.9 & $1.45 \pm 0.02$ & $(4.38 \pm 0.69) \times 10^{29}$ & n & y & y \\                                                                  
GRB090418A & 1.61 & \cite{GCN9151} & 56.3 & $1.27 \pm 0.02$ & $(2.07 \pm 0.22) \times 10^{30}$ & n & y & n \\                                                                   
GRB090423  & 8.26 & \cite{GCN9219} & 10.3 & $1.37 \pm 0.04$ & $(2.28 \pm 0.32) \times 10^{30}$ & n & y & y \\                                                                   
GRB090424  & 0.54 & \cite{GCN9243} & 49.5 & $1.11 \pm 0.01$ & $(1.32 \pm 0.10) \times 10^{30}$ & n & y & n \\                                                                   
GRB090426  & 2.61 & \cite{GCN9264} & 1.2 & $1.04 \pm 0.05$ & $(4.63 \pm 0.68) \times 10^{29}$ & n & y & n \\                                                                    
GRB090429B & 9.20 & \cite{2011ApJ...736....7C} & 5.6 & $1.49 \pm 0.13$ & $(3.09 \pm 0.63) \times 10^{30}$ & n & y & n \\                                                        
GRB090510  & 0.90 & \cite{GCN9353} & 5.7 & $1.59 \pm 0.04$ & $(2.39 \pm 0.28) \times 10^{29}$ & n & y & n \\                                                                    
GRB090516A & 4.11 & \cite{GCN9383} & 163.4 & $1.35 \pm 0.02$ & $(8.42 \pm 1.23) \times 10^{30}$ & n & y & y \\                                                                  
GRB090519  & 3.85 & \cite{GCN9409} & 58.0 & $1.23 \pm 0.14$ & $(9.25^{+2.45}_{-2.41}) \times 10^{28}$ & n & y & y \\                                                            
GRB090529A & 2.63 & \cite{GCN9457} & 70.4 & $0.70 \pm 0.04$ & $(1.20 \pm 0.31) \times 10^{29}$ & n & y & n \\                                                                   
GRB090530  & 1.27 & \cite{GCN15571} & 40.5 & $0.81 \pm 0.02$ & $(1.13 \pm 0.17) \times 10^{29}$ & n & y & y \\                                                                  
GRB090618  & 0.54 & \cite{GCN9518} & 113.3 & $1.15 \pm 0.01$ & $(1.14 \pm 0.05) \times 10^{30}$ & n & y & y \\                                                                  
GRB090715B & 3.00 & \cite{GCN9673} & 266.4 & $1.20 \pm 0.07$ & $(6.95 \pm 4.05) \times 10^{28}$ & n & y & y \\                                                                  
GRB090809A & 2.74 & \cite{GCN9761} & 8.9 & $1.20 \pm 0.08$ & $(4.39 \pm 4.24) \times 10^{28}$ & y & y & y \\                                                                    
GRB090812  & 2.45 & \cite{GCN9771} & 74.5 & $1.37 \pm 0.10$ & $(4.37 \pm 0.41) \times 10^{30}$ & n & y & y \\                                                                   
GRB090814A & 0.70 & \cite{GCN9797} & 78.1 & $1.45^{+0.16}_{-0.14}$ & $(7.17 \pm 1.83) \times 10^{28}$ & y & n & n \\                                                            
GRB090926B & 1.24 & \cite{GCN9947} & 99.3 & $1.11 \pm 0.08$ & $(3.89 \pm 0.65) \times 10^{29}$ & y & n & n \\                                                                   
GRB091018  & 0.97 & \cite{GCN10038} & 4.4 & $1.18 \pm 0.01$ & $(7.03 \pm 0.55) \times 10^{29}$ & n & y & n \\                                                                   
GRB091020  & 1.71 & \cite{GCN10053} & 38.9 & $1.20 \pm 0.01$ & $(2.64 \pm 0.18) \times 10^{30}$ & n & y & n \\                                                                  
GRB091029  & 2.75 & \cite{GCN10100} & 39.2 & $0.81 \pm 0.01$ & $(6.90 \pm 0.66) \times 10^{29}$ & n & y & y \\                                                                  
GRB091109A & 3.08 & \cite{GCN10350} & 48.0 & $1.01 \pm 0.06$ & $(4.95 \pm 0.84) \times 10^{29}$ & n & n & n \\                                                                  
GRB091208B & 1.06 & \cite{GCN10272} & 14.8 & $1.06 \pm 0.02$ & $(4.86 \pm 0.66) \times 10^{29}$ & n & y & n \\                                                                  
GRB100219A & 4.67 & \cite{GCN10445} & 22.2 & $1.35 \pm 0.05$ & $(1.06 \pm 0.19) \times 10^{30}$ & n & y & n \\                                                                  
GRB100302A & 4.81 & \cite{GCN10466} & 18.0 & $0.72 \pm 0.03$ & $(2.68 \pm 0.58) \times 10^{29}$ & n & y & y \\                                                                  
GRB100316B & 1.18 & \cite{GCN10495} & 3.9 & $1.09 \pm 0.06$ & $(1.17 \pm 0.33) \times 10^{29}$ & n & y & n \\                                                                   
GRB100316D & 0.06 & \cite{GCN10512} & 518.7 & $1.53 \pm 0.01$ & $(2.31 \pm 1.61) \times 10^{28}$ & n & y & n \\                                                                 
GRB100418A & 0.62 & \cite{GCN10620} & 7.7 & $0.51 \pm 0.03$ & $(1.89 \pm 0.80) \times 10^{27}$ & n & y & y \\                                                                   
GRB100424A & 2.46 & \cite{GCN14291} & 86.2 & $2.24 \pm 0.16$ & $(2.49 \pm 0.48) \times 10^{30}$ & n & y & n \\                                                                  
GRB100425A & 1.75 & \cite{GCN10684} & 39.0 & $0.75 \pm 0.03$ & $(1.17 \pm 0.26) \times 10^{29}$ & n & y & y \\                                                                  
GRB100513A & 4.77 & \cite{GCN10752} & 72.5 & $0.88 \pm 0.04$ & $(1.08 \pm 0.21) \times 10^{30}$ & n & y & y \\                                                                  
GRB100615A & 1.40 & \cite{GCN14264} & 38.8 & $0.73 \pm 0.02$ & $(2.04 \pm 0.30) \times 10^{30}$ & n & y & n \\                                                                  
GRB100621A & 0.54 & \cite{GCN10876} & 63.6 & $0.97 \pm 0.01$ & $(5.13 \pm 0.49) \times 10^{29}$ & n & y & y \\                                                                  
GRB100724A & 1.29 & \cite{GCN10971} & 1.4 & $1.13 \pm 0.05$ & $(1.23 \pm 0.26) \times 10^{29}$ & n & y & n \\                                                                   
GRB100728A & 1.57 & \cite{GCN14500} & 198.6 & $1.30 \pm 0.01$ & $(1.25 \pm 0.17) \times 10^{31}$ & n & y & y \\                                                                 
GRB100728B & 2.11 & \cite{GCN11317} & 12.1 & $1.49 \pm 0.07$ & $(1.15 \pm 0.22) \times 10^{30}$ & n & y & y \\                                                                  
GRB100805A & 1.85 & \cite{2012MNRAS.426L..86O} & 16.7 & $1.00 \pm 0.03$ & $(2.52 \pm 0.46) \times 10^{29}$ & n & y & y \\                                                       
GRB100814A & 1.44 & \cite{GCN11089} & 174.7 & $1.02 \pm 0.01$ & $(1.02 \pm 0.09) \times 10^{30}$ & n & y & y \\                                                                 
GRB100816A & 0.80 & \cite{GCN11123} & 2.6 & $0.99 \pm 0.04$ & $(1.10 \pm 0.20) \times 10^{29}$ & n & n & y \\                                                                   
GRB100901A & 1.41 & \cite{GCN11164} & 436.4 & $1.02 \pm 0.02$ & $(1.18 \pm 0.25) \times 10^{28}$ & y & y & y \\                                                                 
GRB100906A & 1.73 & \cite{GCN11230} & 114.6 & $1.30 \pm 0.02$ & $(2.39 \pm 0.21) \times 10^{30}$ & n & y & y \\                                                                 
GRB101219A & 0.72 & \cite{GCN11518} & 0.8 & $2.08 \pm 0.22$ & $(1.57 \pm 0.42) \times 10^{28}$ & n & y & n \\                                                                   
GRB101219B & 0.55 & \cite{GCN11579} & 44.7 & $0.64 \pm 0.03$ & $(3.49 \pm 1.14) \times 10^{27}$ & y & y & y \\                                                                  
GRB110106B & 0.62 & \cite{GCN11538} & 43.5 & $1.04 \pm 0.02$ & $(5.47 \pm 0.97) \times 10^{28}$ & n & y & y \\                                                                  
GRB110128A & 2.34 & \cite{GCN11607} & 14.2 & $0.70 \pm 0.05$ & $(3.63^{+1.02}_{-0.98}) \times 10^{28}$ & n & n & n \\                                                           
GRB110205A & 2.22 & \cite{GCN11638} & 249.4 & $1.59 \pm 0.04$ & $(1.32 \pm 0.09) \times 10^{31}$ & n & n & y \\                                                                 
GRB110213A & 1.46 & \cite{GCN11708} & 48.0 & $1.31 \pm 0.01$ & $(8.36 \pm 1.56) \times 10^{29}$ & n & y & n \\                                                                  
GRB110422A & 1.77 & \cite{GCN11978} & 25.8 & $1.30 \pm 0.01$ & $(8.41 \pm 0.56) \times 10^{30}$ & n & y & n \\                                                                  
GRB110503A & 1.61 & \cite{GCN11993} & 10.1 & $1.17 \pm 0.01$ & $(3.52 \pm 0.22) \times 10^{30}$ & n & y & n \\                                                                  
GRB110715A & 0.82 & \cite{GCN12164} & 13.0 & $1.00 \pm 0.01$ & $(1.44 \pm 0.14) \times 10^{30}$ & n & y & n \\                                                                  
GRB110726A & 1.04 & \cite{GCN12202} & 5.2 & $1.11 \pm 0.11$ & $(5.55^{+1.17}_{-1.13}) \times 10^{28}$ & n & y & y \\                                                            
GRB110731A & 2.83 & \cite{GCN12225} & 40.9 & $1.21 \pm 0.02$ & $(7.10 \pm 0.70) \times 10^{30}$ & n & y & y \\                                                                  
GRB110801A & 1.86 & \cite{GCN12234} & 385.3 & $1.28 \pm 0.04$ & $(3.02 \pm 0.32) \times 10^{30}$ & n & y & y \\                                                                 
GRB110808A & 1.35 & \cite{GCN12258} & 40.7 & $0.72 \pm 0.03$ & $(4.53 \pm 2.81) \times 10^{28}$ & n & y & n \\                                                                  
GRB110818A & 3.36 & \cite{GCN12284} & 102.8 & $1.19 \pm 0.03$ & $(2.07 \pm 0.23) \times 10^{30}$ & n & y & n \\                                                                 
GRB111008A & 4.99 & \cite{GCN12431} & 65.3 & $1.08 \pm 0.02$ & $(5.87 \pm 0.51) \times 10^{30}$ & n & y & n \\                                                                  
GRB111107A & 2.89 & \cite{GCN12537} & 26.6 & $0.85 \pm 0.05$ & $(5.04^{+0.89}_{-0.86}) \times 10^{29}$ & n & n & y \\                                                           
GRB111123A & 3.15 & \cite{GCN14273} & 290.0 & $1.14 \pm 0.03$ & $(1.21 \pm 0.23) \times 10^{31}$ & n & y & y \\                                                                 
GRB111209A & 0.68 & \cite{GCN12648} & 733.4 & $1.37 \pm 0.04$ & $(3.95 \pm 0.76) \times 10^{30}$ & y & n & y \\                                                                 
GRB111215A & 2.06 & \cite{2015MNRAS.446.4116V} & 373.8 & $1.34 \pm 0.01$ & $(1.37 \pm 0.14) \times 10^{31}$ & n & y & y \\                                                      
GRB111225A & 0.30 & \cite{GCN16079} & 105.7 & $0.94 \pm 0.14$ & $(2.52 \pm 2.03) \times 10^{24}$ & y & y & y \\                                                                 
GRB111228A & 0.71 & \cite{GCN12761} & 101.2 & $0.90 \pm 0.01$ & $(1.20 \pm 0.10) \times 10^{29}$ & n & y & n \\                                                                 
GRB111229A & 1.38 & \cite{GCN12777} & 439.1 & $0.59 \pm 0.05$ & $(6.88 \pm 1.18) \times 10^{28}$ & n & y & n \\                                                                 
GRB120118B & 2.94 & \cite{GCN14225} & 54.7 & $1.04 \pm 0.08$ & $(3.58 \pm 2.49) \times 10^{29}$ & n & y & y \\                                                                  
GRB120119A & 1.73 & \cite{GCN12865} & 68.0 & $1.40 \pm 0.03$ & $(1.67 \pm 0.19) \times 10^{30}$ & n & y & n \\                                                                  
GRB120326A & 1.80 & \cite{GCN13118} & 69.6 & $0.77 \pm 0.03$ & $(5.43 \pm 1.11) \times 10^{28}$ & n & y & y \\                                                                  
GRB120327A & 2.81 & \cite{GCN13134} & 63.5 & $1.25 \pm 0.03$ & $(2.18^{+0.36}_{-0.33}) \times 10^{30}$ & n & y & y \\                                                           
GRB120404A & 2.88 & \cite{GCN13217} & 37.2 & $1.00 \pm 0.05$ & $(1.12 \pm 0.14) \times 10^{30}$ & n & y & y \\                                                                  
GRB120422A & 0.28 & \cite{GCN13257} & 60.4 & $0.45 \pm 0.03$ & $(2.62 \pm 0.95) \times 10^{26}$ & n & y & n \\                                                                  
GRB120521C & 6.00 & \cite{GCN13348} & 26.7 & $0.71 \pm 0.07$ & $(3.56^{+0.91}_{-0.89}) \times 10^{29}$ & n & y & n \\                                                           
GRB120712A & 4.17 & \cite{GCN13460} & 14.8 & $1.31 \pm 0.05$ & $(4.67 \pm 0.55) \times 10^{30}$ & n & n & n \\                                                                  
GRB120724A & 1.48 & \cite{GCN13512} & 77.9 & $0.47 \pm 0.09$ & $(1.84 \pm 0.61) \times 10^{28}$ & n & n & y \\                                                                  
GRB120729A & 0.80 & \cite{GCN13532} & 93.9 & $1.60 \pm 0.03$ & $(4.04 \pm 0.44) \times 10^{29}$ & n & y & n \\                                                                  
GRB120802A & 3.80 & \cite{GCN13562} & 50.3 & $0.62 \pm 0.13$ & $(5.55 \pm 1.14) \times 10^{29}$ & n & y & n \\                                                                  
GRB120811C & 2.67 & \cite{GCN13628} & 26.8 & $0.98 \pm 0.03$ & $(1.64 \pm 0.20) \times 10^{30}$ & n & y & n \\                                                                  
GRB120907A & 0.97 & \cite{GCN13723} & 6.1 & $0.94 \pm 0.02$ & $(1.39 \pm 0.15) \times 10^{29}$ & n & y & n \\                                                                   
GRB120922A & 3.10 & \cite{GCN13810} & 161.4 & $0.98 \pm 0.02$ & $(2.90 \pm 0.61) \times 10^{30}$ & n & y & y \\                                                                 
GRB121024A & 2.30 & \cite{GCN13890} & 68.0 & $1.17 \pm 0.04$ & $(1.76 \pm 0.23) \times 10^{30}$ & n & y & y \\                                                                  
GRB121027A & 1.77 & \cite{2014ApJ...781...13L} & 95.7 & $0.75 \pm 0.02$ & $(4.45 \pm 0.75) \times 10^{29}$ & n & y & y \\                                                       
GRB121128A & 2.20 & \cite{GCN14009} & 23.4 & $1.46 \pm 0.03$ & $(5.10 \pm 0.41) \times 10^{30}$ & n & y & y \\                                                                  
GRB121201A & 3.38 & \cite{GCN14035} & 38.0 & $1.15 \pm 0.06$ & $(3.43 \pm 0.54) \times 10^{29}$ & n & n & n \\                                                                  
GRB121211A & 1.02 & \cite{GCN14059} & 182.7 & $0.98 \pm 0.03$ & $(5.32 \pm 0.74) \times 10^{29}$ & n & n & y \\                                                                 
GRB121229A & 2.71 & \cite{GCN14120} & 86.5 & $0.69 \pm 0.15$ & $(2.71 \pm 1.03) \times 10^{29}$ & y & n & y \\                                                                  
GRB130131B & 2.54 & \cite{GCN14286} & 4.3 & $1.24 \pm 0.27$ & $(1.04 \pm 0.26) \times 10^{30}$ & n & n & y \\                                                                   
GRB130408A & 3.76 & \cite{GCN14365} & 28.7 & $1.59 \pm 0.06$ & $(7.17 \pm 0.74) \times 10^{30}$ & n & y & y \\                                                                  
GRB130418A & 1.22 & \cite{GCN14380} & 274.9 & $1.47 \pm 0.06$ & $(1.25 \pm 0.24) \times 10^{30}$ & n & y & n \\                                                                 
GRB130420A & 1.30 & \cite{GCN14437} & 121.1 & $0.89 \pm 0.01$ & $(3.80 \pm 0.46) \times 10^{29}$ & n & y & n \\                                                                 
GRB130427A & 0.34 & \cite{GCN14455} & 241.3 & $1.28 \pm 0.01$ & $(8.16 \pm 1.23) \times 10^{30}$ & n & y & n \\                                                                 
GRB130427B & 2.78 & \cite{GCN14493} & 25.9 & $1.24 \pm 0.05$ & $(7.38^{+1.45}_{-1.35}) \times 10^{29}$ & n & y & y \\                                                           
GRB130505A & 2.27 & \cite{GCN14567} & 88.5 & $1.38 \pm 0.01$ & $(2.74 \pm 0.15) \times 10^{31}$ & n & y & n \\                                                                  
GRB130511A & 1.30 & \cite{GCN14621} & 5.4 & $1.20 \pm 0.05$ & $(1.88 \pm 0.32) \times 10^{29}$ & n & y & n \\                                                                   
GRB130514A & 3.60 & \cite{GCN14634} & 191.7 & $1.02 \pm 0.03$ & $(6.48 \pm 1.02) \times 10^{30}$ & y & n & y \\                                                                 
GRB130603B & 0.36 & \cite{GCN14744} & 0.2 & $1.14 \pm 0.03$ & $(2.71 \pm 0.34) \times 10^{28}$ & n & y & n \\                                                                   
GRB130604A & 1.06 & \cite{GCN14762} & 76.3 & $0.28 \pm 0.29$ & $(5.88 \pm 1.50) \times 10^{29}$ & n & n & y \\                                                                  
GRB130606A & 5.91 & \cite{GCN14796} & 262.3 & $1.35 \pm 0.04$ & $(7.27 \pm 0.78) \times 10^{30}$ & n & n & y \\                                                                 
GRB130610A & 2.09 & \cite{GCN14848} & 61.5 & $1.20 \pm 0.04$ & $(7.32 \pm 0.96) \times 10^{29}$ & n & n & n \\                                                                  
GRB130612A & 2.01 & \cite{GCN14882} & 40.5 & $0.94 \pm 0.05$ & $(1.55 \pm 0.53) \times 10^{29}$ & n & y & y \\                                                                  
GRB130701A & 1.15 & \cite{GCN14956} & 4.4 & $1.25 \pm 0.03$ & $(1.39 \pm 0.18) \times 10^{30}$ & n & y & n \\                                                                   
GRB130831A & 0.48 & \cite{GCN15144} & 30.2 & $1.12 \pm 0.02$ & $(1.01 \pm 0.23) \times 10^{29}$ & n & y & y \\                                                                  
GRB130907A & 1.24 & \cite{GCN15187} & 359.0 & $1.44 \pm 0.12$ & $(3.39 \pm 0.19) \times 10^{31}$ & n & n & y \\                                                                 
GRB130925A & 0.35 & \cite{GCN15249} & 161.1 & $1.05 \pm 0.01$ & $(2.31 \pm 0.46) \times 10^{30}$ & n & y & y \\                                                                 
GRB131004A & 0.72 & \cite{GCN15307} & 1.5 & $1.20^{+0.21}_{-0.17}$ & $(9.96 \pm 2.07) \times 10^{28}$ & n & y & y \\                                                            
GRB131030A & 1.29 & \cite{GCN15407} & 39.4 & $1.12 \pm 0.01$ & $(3.99 \pm 0.54) \times 10^{30}$ & n & y & y \\                                                                  
GRB131103A & 0.60 & \cite{GCN15451} & 14.7 & $1.06 \pm 0.04$ & $(7.07 \pm 1.56) \times 10^{28}$ & n & y & y \\                                                                  
GRB131105A & 1.69 & \cite{GCN15450} & 111.8 & $0.84 \pm 0.02$ & $(4.92 \pm 0.67) \times 10^{29}$ & n & y & n \\                                                                 
GRB131117A & 4.04 & \cite{GCN15494} & 10.9 & $1.03 \pm 0.07$ & $(1.04 \pm 0.19) \times 10^{30}$ & n & y & y \\                                                                  
GRB131227A & 5.30 & \cite{GCN15624} & 18.0 & $1.36 \pm 0.18$ & $(4.74 \pm 0.98) \times 10^{30}$ & n & y & n \\                                                                  
GRB140206A & 2.73 & \cite{GCN15802} & 93.6 & $1.05 \pm 0.01$ & $(1.49 \pm 0.08) \times 10^{31}$ & n & y & y \\                                                                  
GRB140301A & 1.42 & \cite{GCN15900} & 36.0 & $1.14^{+0.11}_{-0.13}$ & $(4.48 \pm 1.08) \times 10^{29}$ & n & y & y \\                                                           
GRB140304A & 5.30 & \cite{GCN15936} & 14.8 & $2.31 \pm 0.12$ & $(2.07 \pm 0.34) \times 10^{31}$ & n & y & y \\                                                                  
GRB140318A & 1.02 & \cite{GCN15988} & 7.6 & $0.95 \pm 0.08$ & $(1.30 \pm 0.38) \times 10^{30}$ & n & n & n \\                                                                   